\documentclass[a4paper,fleqn,usenatbib]{mnras}

\usepackage[T1]{fontenc}
\usepackage{ae,aecompl}
\usepackage{empheq}
\usepackage{xcolor}
\colorlet{RED}{red}

\usepackage{graphicx}	
\usepackage{amsmath}	
\usepackage{amssymb}	

\usepackage{amsfonts}
\usepackage{mathrsfs} 

\usepackage{subcaption}
\def\beq{\begin{equation}}
\def\eeq{\end{equation}}
\def\bea{\begin{eqnarray}}
\def\eea{\end{eqnarray}}
\def\bi{\begin{itemize}}
\def\ei{\end{itemize}}
\def\ev{\, {\rm eV}}
\def\neff{N_{\rm eff}}
\def\lcdm{{$\Lambda$CDM }}
\def\camb{{\sc camb }}
\def\planck{{\it Planck}}
\def\mpch{\,h^{-1}\,{\rm Mpc}}
\def\msol{M_\odot}

\newcommand{\vc}[1]{\textit{\textbf{#1}}}  

\title[Dynamical friction and neutrinos]{Dynamical friction in the primordial neutrino sea}

\author[Okoli et al. 2017]{Chiamaka Okoli$^{1,2}$\thanks{E-mail: c2okoli@uwaterloo.ca}, Morag I. Scrimgeour$^{1}$, Niayesh Afshordi$^{1,2}$\thanks{E-mail: nafshordi@pitp.ca}, and \newauthor{Michael J. Hudson$^{1,2}$\thanks{E-mail: mike.hudson@uwaterloo.ca}}\\
$^{1}$Department of Physics and Astronomy, University of Waterloo, 200 University Avenue West, Waterloo,ON  N2L 3G1, Canada\\
$^{2}$Perimeter Institute for Theoretical Physics, 31 Caroline Street North, Waterloo, ON N2L 2Y5, Canada}

\date{Accepted XXX. Received YYY; in original form ZZZ}

\pubyear{2017}

\begin{document}

\label{firstpage}
\pagerange{\pageref{firstpage}--\pageref{lastpage}}
\maketitle




\begin{abstract}
Standard big bang cosmology predicts a cosmic neutrino background at $T_\nu \simeq 1.95~$K. Given the current neutrino oscillation measurements, we know most neutrinos move at large, but non-relativistic, velocities. Therefore, dark matter haloes moving in the sea of primordial neutrinos form a neutrino wake behind them, which would slow them down, due to the effect of {\it dynamical friction}. In this paper, we quantify this effect for realistic haloes, in the context of the halo model of structure formation, and show that it scales as $m_\nu^4$ $\times$ relative velocity, and monotonically grows with the halo mass.  Galaxy redshift surveys can be sensitive to this effect (at $>3\sigma$ confidence level, depending on survey properties, neutrino mass and hierarchy) through redshift space distortions (RSD) of distinct galaxy populations.

\end{abstract}

\begin{keywords}
dark matter: cosmology --- haloes: galaxies --- theory: cosmology
\end{keywords}

\section{Introduction}
Neutrinos are an interesting component of the standard model of particles that play a number of unique roles in particle physics, astroparticle physics and cosmology. In the Standard Model of particle physics, they are depicted as massless and not clearly associated as Majorana or Dirac particles. Observations of flavour oscillations in solar and atmospheric neutrinos point to the existence of neutrino mass. In cosmology, the presence of massive active neutrinos potentially resolves some of the tension between the observed galaxy counts and those predicted by the {\it Planck} cosmological parameters as suggested by \cite{mark2014,battye2014,beutler2014}.

Constraints on the mass-squared difference of neutrinos via oscillation experiments give best-fit values of $\Delta m_{12}^2  \equiv m_2^2 - m_1^2 = 7.54 \times 10^{-5} {\rm eV}^2$ and $\Delta m_{13}^2 \equiv m_3^2 - (m_1^2 + m_2^2)/2 = 2.43(2) \times 10^{-3} {\rm eV}^2$ for the normal (inverted) hierarchy \citep{fogli,malt}. These constraints lead to a lower limit on the sum of the masses of these neutrinos, $M_{\nu} \equiv \sum_i m_i > 0.058 \,{\rm eV}$ for the normal hierarchy and $M_{\nu}  > 0.11 \, {\rm eV}$ for the inverted hierarchy \citep[see][for a more comprehensive review of the role of massive neutrinos in cosmology]{les,les12}. Thus, any limit on $M_{\nu} < 0.1 \,{\rm eV}$ rules out the inverted neutrino hierarchy. Alternative limits on the sum of the neutrino mass $M_{\nu}$ may also be placed by cosmological observations and measurements. Although some degeneracy exists between the Hubble constant and the neutrino mass on the background cosmology, the cosmic microwave background (CMB) temperature perturbations are affected by the neutrino mass through the early-time \textcolor{black}{integrated Sachs-Wolfe (}ISW\textcolor{black}{)} effect and the lensing effect on the power spectrum. The baryon acoustic oscillations (BAO) and measurements of the CMB temperature anisotropy have placed limits on the sum of the mass of the neutrinos as $M_{\nu} < 0.23 \,{\rm eV}$ \citep{planck}; albeit, slightly dependent on the assumed cosmological model parameters. \textcolor{black}{First described by \citet{1980PhRvL..45.1980B},} galaxy surveys present yet another method to constrain the mass of neutrinos as presented in \cite{riemer}. Suppression of structure below the free-streaming scale of the neutrinos, when they first become non-relativistic, leads to a decline in the matter power spectrum (usually at the per cent level). However, in practice, galaxy power spectra are measured and so precise knowledge of the galaxy bias as a function of scale is required since the galaxies do not cluster as matter. Notwithstanding, \cite{riemer} explored this technique in the WiggleZ Dark Energy Survey and placed an upper limit on the sum of the neutrino masses, \textcolor{black}{$M_{\nu} < 0.18 \,{\rm eV}$ for three degenerate neutrino species with no prior placed on the minimum sum of neutrino masses}. More recent estimates by \cite{cuesta2015} using the WiggleZ  Dark Energy Survey  and SDSS-DR7 LRG, together with the BAO and CMB temperature and polarisation anisotropies measurements by \textit{Planck} have yielded even tighter constraints of $M_{\nu} < 0.13 \,{\rm eV}$  at the $95\%$ C.L. on the sum of the neutrino masses . 

\begin{figure}
\begin{center}
\includegraphics[width=9cm]{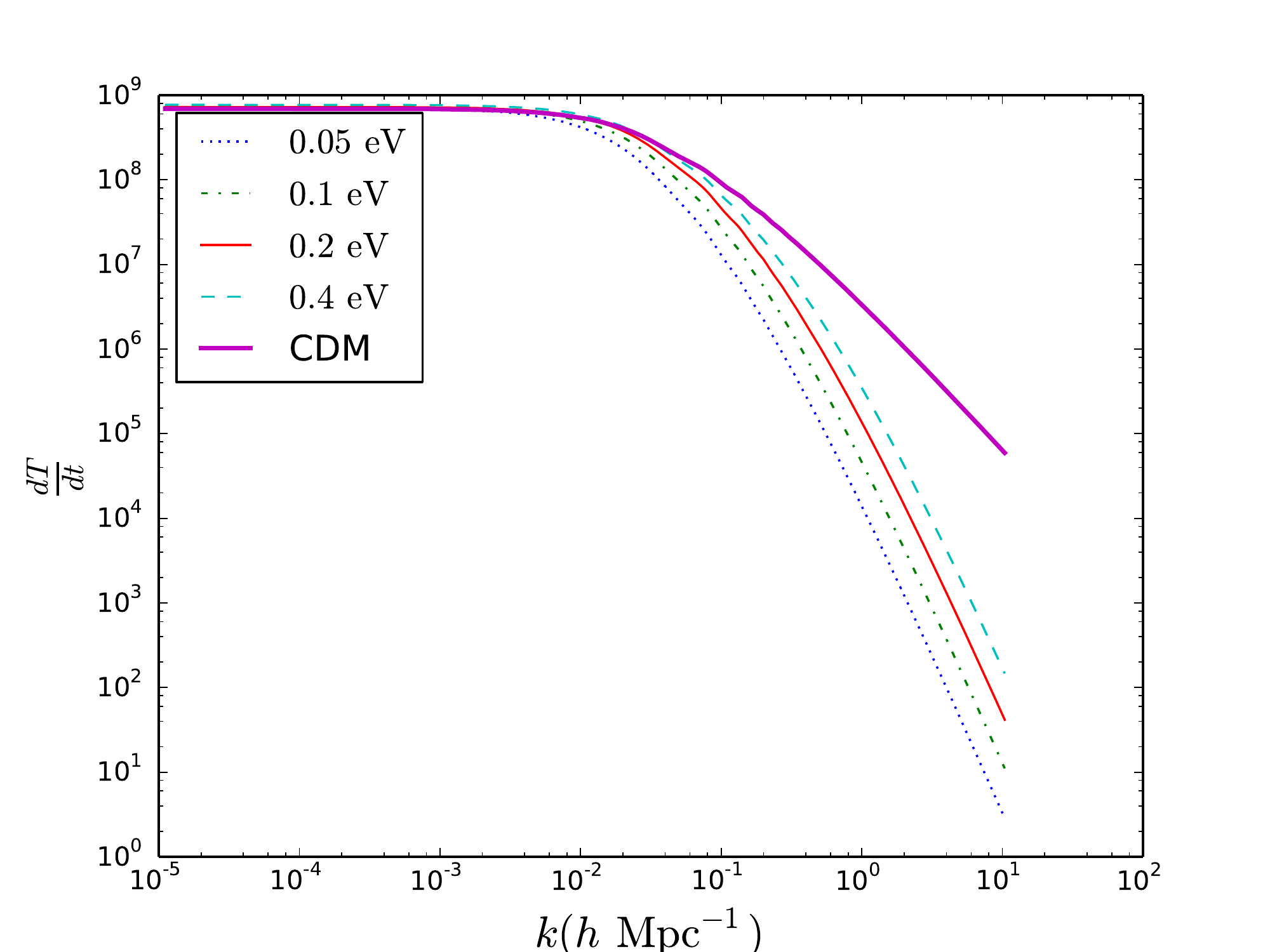}
\caption{Derivative of the transfer function, $\dot{T}(k) = dT(k)/dt$, at $z=0$ for cold dark matter and neutrinos of different masses.} 
\label{fig:dot_T}
\end{center}
\end{figure}
\section{The Neutrino - Cold Dark Matter relative velocity}
\label{sec:nucdm}

A complementary technique to measure the neutrino masses using their peculiar velocities relative to dark matter has been suggested by \cite{zhu}. \textcolor{black}{The relative velocity between the neutrinos and dark matter leads to an observable dipole distortion in the cross correlation of different tracers.} This effect is similar to the relative velocity between baryons and cold dark matter, first suggested by \cite{tseliakhovich2010} and explored in a number of other works, including \cite{yoo2013}. The neutrino particles stream coherently across the \textcolor{black}{cold dark matter (}CDM\textcolor{black}{)} haloes over a coherence/Jeans scale of $20-50~ h^{-1}$Mpc. 

In this work, we provide a semi-analytic derivation of this effect in the nonlinear regime, which is equivalent to the dynamical friction for CDM haloes moving in the primordial neutrino sea. The outline is as follows: Section \ref{sec:nucdm} introduces the relations between velocities and density perturbations for both neutrinos and cold dark matter in the linear regime. Section \ref{sec:dynafriction} examines the effect of dynamical friction on the dark matter structures due to the streaming neutrinos. We then present the methodology, and expected signal to noise for detecting this effect in current and  future surveys in Section \ref{sec:sigtonoise}. Finally,  Section \ref{sec:discussion} summarizes our results and concludes the paper, while the appendices provide details of signal-to-noise calculations, and comparison to nonlinear effects in $\Lambda$CDM simulations without neutrinos. 

For the calculations below, we use a \lcdm \camb \citep{lewis2000} power spectrum with \planck\footnote{\cite{planck13}. We use the Planck-only best fit values.} parameters $(\Omega_{m0},\Omega_{\Lambda0}, \Omega_{0b}) = (0.32,0.68,0.049)$, Hubble parameter $h=0.67$, rms \textcolor{black}{(root mean square)} density fluctuation in $8\mpch$ spheres $\sigma_8 = 0.8344$ and scalar spectral index $n_s=0.963$. We also utilize the analytic power spectrum of \cite{eisenstein1998} where necessary. The dark energy density and total matter density $\Omega_{m0}= \Omega_{c0} + \Omega_{b0} + \Omega_{\nu 0}$ are kept fixed while the cold dark matter density $\Omega_{c0}$ and the neutrino density $\Omega_{\nu 0}$ are adjusted as needed. The relation between the neutrino density and neutrino mass is given by 
\begin{equation}
\Omega_{\nu0} = \frac{\sum_{i=1}^3  m_{i} }{94.07 h^2 \ev},
\end{equation}
with the sum being over the three neutrino species. We assume an effective number of relativistic species $\neff = 3.046$ \textcolor{black}{\citep{planck}}. The notation $M_{\nu} \equiv \sum_{i=1}^3  m_{i}$ will be mostly used to replace the sum over neutrino mass in the rest of this article.

\begin{figure}
\begin{center}
\includegraphics[width=9cm]{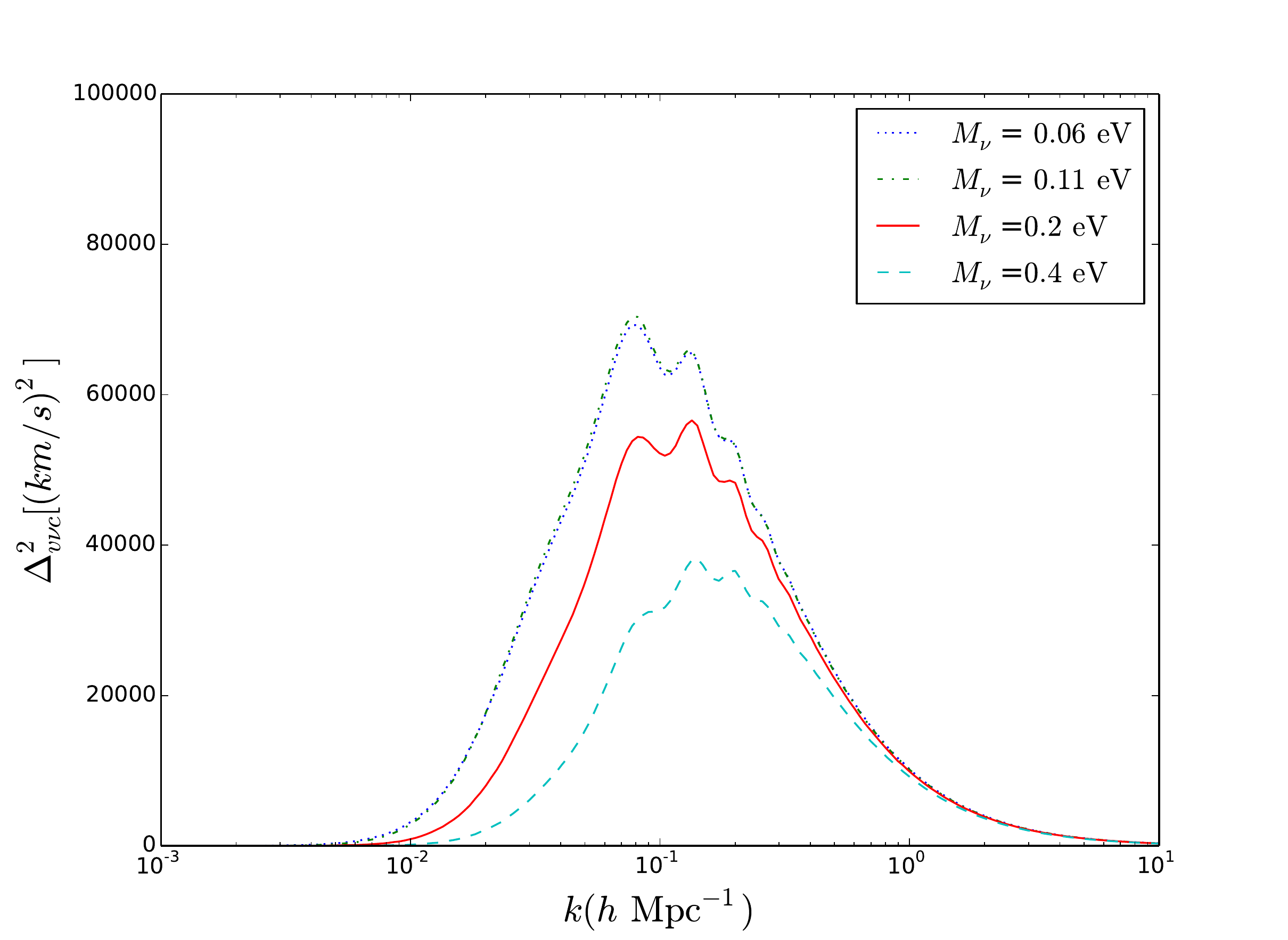}
\caption{The power spectrum of the relative velocity between the neutrinos and cold dark matter. The different curves represent different values of $M_\nu = \sum  m_{\nu}$ (assuming normal hierarchy). \textcolor{black}{The two lowest $M_\nu$'s are close due to the fact that the most massive of the three neutrinos in each sum have similar masses and dominate in the sum.} } 
\label{Delta2_vnc}
\end{center}
\end{figure}

We proceed to calculate the CDM-neutrino relative velocity power spectrum following \cite{tseliakhovich2010} and \cite{zhu}. The relative velocity, as we will see in the next section, is relevant for estimating the dynamical friction on a halo due to neutrinos. Assuming the linearized continuity equation holds for both neutrinos and cold dark matter \textcolor{black}{(on scales $k \lesssim 1\rm{h/Mpc}$)}, in Fourier space it is given by 
\begin{equation}
\label{cont_eqn_k}
\vc{v}(k) = - \frac{i \vc{k} a}{k^2} \dot{\delta}(k)
\end{equation}
where $a=1/(1+z)$ is the cosmic scale factor and we have assumed that there is no vorticity (curl of the vector perturbation is zero). This is tested in \cite{inman2015}, where the velocity was seen to be curl-free on scales $k \lesssim 1\rm{h/Mpc}.$

The variance in the relative velocity ${\vc{v}_{\nu c} \equiv \vc{v}_\nu - \vc{v}_c}$ between neutrinos ($\nu$) and cold dark matter ($c$) can be calculated to be
\begin{eqnarray}
\label{v_nuc2}
\langle v^2_{\nu c}(R,a) \rangle &=&  \frac{a^2}{2\pi^2} \int k^2 dk    \left[ \frac{\dot{\delta}_\nu({k,a}) - \dot{\delta}_c({k,a})}{k} \right]^2 \widetilde{W}^2(kR) \nonumber \\
&=&   a^2\int  \frac{dk}{k} \Delta^2_{v \nu c}(k,z) \widetilde{W}^2(kR)
\end{eqnarray}
where $\widetilde{W}(kR)$ \textcolor{black}{is the Fourier transform of the }window function, and $\Delta^2_{v \nu c}(k,z)$ is the neutrino - cold dark matter relative velocity power spectrum. \textcolor{black}{The essence of the window function is to filter the velocity perturbation field to get a smooth field. We will adopt one of the most commonly used window functions -- the spherical top-hat \footnote{\textcolor{black}{The advantage of using top-hat window function is that it is localized in real space, and thus can be applied to finite real-space data. However, we do not expect our conclusions to be sensitive to this choice.}}-- defined  as 
\begin{equation}
W(x;R) = \left(\frac{4 \pi}{3} R^3\right)^{-1}
\begin{cases}
\, 1 \qquad \mbox{for} \, |x| \leq R \\ 
\, 0 \qquad \mbox{for} \, |x| \geq R.
\end{cases}
\end{equation}
In Fourier space, it is given by
\begin{equation}
\widetilde{W}(kR) = \frac{ 3 j_1(kR)}{kR},
\end{equation}
where $j_1$ is the first order spherical Bessel function.} \textcolor{black}{The relative velocity power spectrum} can be written in terms of the transfer functions of neutrinos and CDM as
\begin{equation}
 \Delta^2_{v \nu c}(k,z) = \mathcal{P}_\chi \left[ \frac{\dot{T}_\nu({k,z}) - \dot{T}_c({k,z})}{k} \right]^2 \label{Delta_v},
\end{equation}
where $\mathcal{P}_\chi \propto k^{n_s +3}/(2\pi^2)$ is the primordial power spectrum of density perturbations and $\dot{T}(k,z)$ is the time derivative of the transfer function at redshift $z$.

We show the derivative of the transfer function, $\dot{T}$, at $z=0$ for cold dark matter and neutrinos of different masses, in Figure \ref{fig:dot_T}\textcolor{black}{, which demonstrates the impact of free steaming  on their density perturbations. As expected, more free streaming leads to more suppression of growth on small scales for lighter neutrinos.} The relative velocity power spectrum for different sum of neutrino masses (unless otherwise stated, all sum of neutrino masses assume the normal hierarchy) is shown in Figure \ref{Delta2_vnc}.

Obviously in Figure~\ref{Delta2_vnc}, the power lies in the range $k\sim[0.01,1]$. The \textcolor{black}{rms} relative velocity $v_{\nu c}$ within a sphere as a function of the radius of the window function used is shown in Figure ~\ref{v_nuc_r} for different sum of neutrino masses. \textcolor{black}{The relative velocity between the neutrino and dark matter decreases as the mass in the neutrinos increases.}

\section{Dynamical friction due to massive neutrinos}
\label{sec:dynafriction}
Dark matter haloes sitting in a streaming background of neutrinos will experience a deceleration due to dynamical friction. Larger haloes experience a larger dynamical friction force than smaller-mass haloes, and so larger- and smaller- mass haloes will have different displacements relative to the neutrino streaming direction.

In this section, we calculate the general structure of massive neutrino wakes, and their dynamical friction force on dark matter haloes with a Navarro-Frenk-White (NFW) profile \citep{navarro1996}. This drag on the halo should lead to a displacement, which is nonexistent in the absence of neutrinos. This displacement $\Delta x$, in the halo's position, will be first estimated by approximating dark matter haloes as single spheres in Section~\ref{section_sphereapprox}. We then develop the general formalism for computing the neutrino wake in Fourier space in Section \ref{section_general},  and use it to derive dynamical friction assuming the full halo model in Section~\ref{section_halomodel}. The halo model consists of contributions from the one-halo and two-halo terms. We shall see that, while the one-halo term is equivalent to the solid sphere approximation, the two-halo term dominates the drag on small haloes. 
\subsection{Solid sphere approximation}
\label{section_sphereapprox}

The phase space distribution of neutrinos is given by the Fermi-Dirac distribution:
\begin{equation}
d\mathcal{N}_\nu = \frac{d^3xd^3p}{(2\pi\hbar)^3} f_\nu(\vc{p}) 
= \frac{d^3xd^3p}{(2\pi\hbar)^3} \frac{2 N_\nu}{\exp(pc/T_\nu)+1},
\end{equation}
where
\begin{equation}
T_\nu = \left( \frac{4}{11} \right)^{1/3} T_{\rm CMB} \simeq 1.95 \,{\rm K},
\end{equation}
\textcolor{black}{$T_{\rm CMB}$ is the temperature of the CMB today, and $N_{\nu}$ is the number of massive neutrino species.}
In the rest frame of a dark matter halo that moves with velocity $\vc{v}_{\nu c}$ relative to the neutrinos, the phase space density takes the form:
\begin{eqnarray}
f_\nu (\vc{p}) &=& \frac{2N_\nu}{\exp(|\vc{p} + m_\nu \vc{v}_{\nu c}|c/T_\nu)+1} \\
&\simeq& \frac{2 N_\nu}{\exp(pc/T_\nu)+1} - \frac{N_\nu m_\nu c \, \vc{v}_{\nu c}\cdot\vc{p}}{pT_\nu [1 + \cosh (pc/T_\nu)]} + \mathcal{O}(v_{\nu c}^2), \nonumber 
\end{eqnarray}
\textcolor{black}{with the factor of 2 accounting} for antineutrinos along with neutrinos. The dynamical friction force is then given by integrating over \textcolor{black}{all range of impact parameter, $b$ and over} a dark matter halo \textcolor{black}{of mass, say $M_h$}:
\begin{eqnarray}
\vc{F}&=& \int (2\pi b db) \frac{2[GM_h(<b)]^2m_\nu^3}{b^2} \int \frac{d^3p}{(2 \pi \hbar)^3} f_\nu (\vc{p})\frac{\vc{p}}{p^3} \nonumber \\
&=& - \frac{2 N_\nu m_\nu^4 \vc{v}_{\nu c}}{3\pi\hbar^3} \int\frac{db}{b} [GM_h(<b)]^2 \nonumber \\
&\simeq& - \frac{2 N_\nu \ln(\Lambda)(GM_h)^2 m_\nu^4 \vc{v}_{\nu c}}{3 \pi \hbar^3}. \label{force_solid}
\end{eqnarray}
Notice that the terms in the integral are similar to the the well-known \textcolor{black}{Chandrasekhar} dynamical friction formula \textcolor{black}{\citep{binney2008}}. The first term in the first integral accounts for the range of impact parameters of interaction between the neutrino and the halo. The second term in the first integral is the drag force due to an interaction between a neutrino a halo. The second integral incorporates the density distribution of the neutrinos. Using Newton's $2^{\rm nd}$ law, $\vc{F} = M \dot{\vc v}$, we can find the change in relative velocity due to dynamical friction, over a Hubble time $t \simeq H^{-1}$:
\begin{figure}
\begin{center}
\includegraphics[width=9cm]{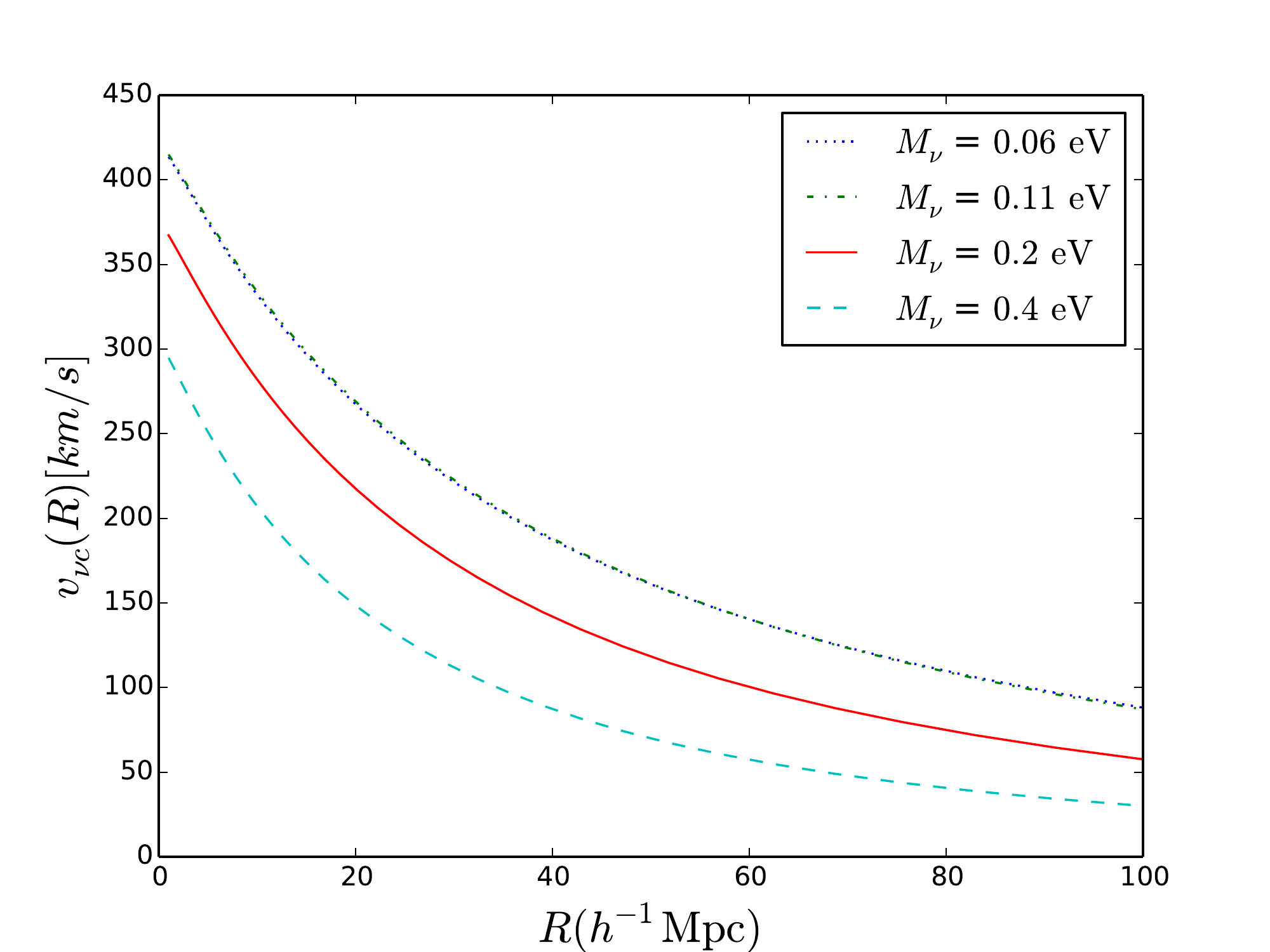}
\caption{Relative CDM-neutrino velocity $v_{\nu c}$ as a function of top hat window function radius $R$, for four different neutrino masses.} 
\label{v_nuc_r}
\end{center}
\end{figure}
\begin{eqnarray}
\label{Delta_u}
\frac{\Delta v_{\nu c}}{v_{\nu c}} &=& \frac{2 N_\nu \ln(\Lambda)G^2M_hm^4_\nu}{3 \pi H \hbar^3} \nonumber \\
&=& 5.6 \times 10^{-3} \, h^{-2} \left(\ln\Lambda \over \ln30 \right) \left(M_h \over 10^{15} h^{-1}  M_{\odot}\right)  \nonumber \\
&& \left( N_{\nu} \over 3\right) \left(m_{\nu} \over 0.1~ {\rm eV} \right)^4, \label{solid}
\end{eqnarray}
where $H=H(z)$ is the Hubble constant, $h$ is the Hubble parameter, and $\ln \Lambda = \int db/b$ is the Coulomb logarithm. \textcolor{black}{$\ln \Lambda \equiv \ln \frac{b_{max}}{b_{min}}$, where $b_{max}$ and $b_{min}$ are the maximum and minimum impact parameters respectively. Here we use  $b_{max} \sim 30$ Mpc as the typical neutrino free steaming length, and  $b_{min} \sim 1$ Mpc as the typical size of CDM haloes.} This leads to an average displacement over a Hubble time of:
 \begin{eqnarray}
 \Delta x &=& \frac{1}{2} \Delta v_{\nu c} \times t \nonumber \\
 &=& 6.66 \, h^{-3} \, {\rm kpc} \left( v_{\nu c} \over 236 ~{\rm km/s}\right) \left(\ln\Lambda \over \ln30 \right) \nonumber \\
 && \left(M_{200} \over 10^{15} h^{-1}M_{\odot}\right) \left( N_{\nu} \over 3\right) \left(m_{\nu} \over 0.1~ {\rm eV} \right)^4.
\end{eqnarray}
If there exist different neutrino mass eigenstates, these will have different velocities and different displacements, which can be calculated independently. The total displacement is equal to the sum of displacements due to each neutrino species $i$, so that $\Delta x_{\rm tot} = \sum_i \Delta x_i$.

\subsection{Dynamical friction: general formalism}
\label{section_general}

Now, let's consider a more realistic dark matter distribution. For large thermal velocity of neutrinos $\frac{T_\nu}{m_\nu c} \gg v_{\rm halo}$, we can assume a steady state neutrino distribution, i.e. it satisfies the time-independent Boltzmann equation:
\beq
\frac{p^i}{m_\nu} \frac{\partial f_\nu({\bf x, p})}{\partial x^i} -m_\nu \nabla_i \Phi({\bf x})  \frac{\partial f_\nu({\bf x, p})}{\partial p^i} =0,
\eeq
where $\Phi({\bf x})$ is the gravitational potential of CDM structure. 
This approximation is also valid in comoving phase space on large scales, as long as the neutrino thermal velocity exceeds the Hubble flow, i.e. $\Delta x \lesssim 25~  {\rm Mpc} \left(m_\nu/{\rm 0.1 ~eV}\right)^{-1}$.

To linear order in $\Phi$ \textcolor{black}{(in line with the assumption of linear regime)}, we can consider linear perturbations $\delta f({\bf x, p}) = f({\bf x, p})  - \bar{f}({\bf p}) $ in Fourier space:
\beq
({\bf p\cdot k}) \delta f_{\nu,{\bf k}}({\bf p}) = m^2_\nu \Phi_{\bf k}\left({\bf k \cdot} \frac{\partial \bar{f}({\bf p}) }{\partial {\bf p}}\right).
\eeq 
The density of the neutrinos in Fourier space is:
\bea
\delta \rho_{\nu, {\bf k}} &=& m_\nu\int \frac{d^3p}{(2\pi\hbar)^3}~ \delta f_{\nu,{\bf k}}({\bf p}) \nonumber \\
&=& m^3_\nu \Phi_{\bf k} k^i \int \frac{d^3p}{(2\pi\hbar)^3}  \frac{1}{  {\bf p\cdot k}} \frac{\partial \bar{f}({\bf p}) }{\partial p^i} \nonumber \\
&=&  2N_\nu m^3_\nu \Phi_{\bf k} k^i  \int \frac{d^3p}{(2\pi\hbar)^3} \nonumber \\
  &&\left[\frac{1}{  ({\bf p}- m_\nu \vc{v}_{\nu c})\cdot {\bf k}} \frac{\partial \left[\exp(|{\bf p}|c/T_\nu)+1\right]^{-1}}{\partial p^i}\right],\nonumber\\
\eea
where we have changed the integration variable $ {\bf p} \rightarrow {\bf p}- m_\nu \vc{v}_{\nu c}$ in the last step, and used the shifted Fermi-Dirac distribution. We can write the integral in spherical coordinates (using ${\bf p\cdot k} = p |{\bf k}| \cos\theta $):
\bea
\delta \rho_{\nu, {\bf k}} &=&  2N_\nu m^3_\nu \Phi_{\bf k} |{\bf k}|  \int_0^\infty \frac{p^2dp}{(2\pi)^2\hbar^3}  \frac{\partial \left[\exp(pc/T_\nu)+1\right]^{-1}}{\partial p} \nonumber \\
&& \int_{-1}^1 \frac{\cos\theta \cdot d\cos\theta}{p|{\bf k}| \cos\theta- m_\nu \vc{v}_{\nu c} \cdot {\bf k}}. \label{delta_rho_int}
\eea

The integral over $\cos\theta$ has a singularity and therefore requires regularization. To this end, we need to set the initial conditions upstream in the neutrino flow. Assuming that the gravitational potential of haloes is turned on gradually as $\exp(- \epsilon {\vc{v}_{\nu c} \cdot {\bf x} })$, which is equivalent to taking ${\bf k} \rightarrow {\bf k} +i \epsilon \vc{v}_{\nu c}$ in the $\exp(i {\bf k\cdot x})$ Fourier phase factor, we can use Sokhatsky-Weierstrass Identity:
\begin{align}
  & \frac{1}{p|{\bf k}| \cos\theta- m_\nu \vc{v}_{\nu c}\cdot {\bf k} -i m_\nu \epsilon |\vc{v}_{\nu c}|^2 } =  \nonumber \\
    & {\rm Pr} \frac{1}{p|{\bf k}| \cos\theta- m_\nu \vc{v}_{\nu c}\cdot {\bf k}} - i \pi \delta_D\left(p|{\bf k}| \cos\theta- m_\nu \vc{v}_{\nu c}\cdot {\bf k} \right). 
\end{align}
This regularization ensures that neutrino wakes form {\it behind} the haloes, and is similar to the one used in the derivation of Landau damping in plasmas \citep[e.g.,][]{2000thas.book.....P}.

Substituting into the angular integral in Eq. (\ref{delta_rho_int}) yields:
\bea
\int_{-1}^1 \frac{\cos\theta \cdot d\cos\theta}{p|{\bf k}| \cos\theta- m_\nu \vc{v}_{\nu c}\cdot {\bf k} -i m_\nu \epsilon |\vc{v}_{\nu c}|^2 } = \nonumber \\
\frac{2}{p|{\bf k}| }+\frac{m_\nu \vc{v}_{\nu c}\cdot {\bf k} }{(p|{\bf k}| )^2}\log\left| p|{\bf k}|- m_\nu \vc{v}_{\nu c} \cdot {\bf k} \over  p|{\bf k}|+ m_\nu \vc{v}_{\nu c} \cdot {\bf k} \right| -  \nonumber \\
\frac{i \pi m_\nu \vc{v}_{\nu c} \cdot {\bf k}}{(p|{\bf k}| )^2} \Theta\left(  p|{\bf k}|- m_\nu |\vc{v}_{\nu c}\cdot {\bf k}|\right).
\eea
The real part of this integral is symmetric under $\vc{v}_{\nu c} \rightarrow -\vc{v}_{\nu c}$, and thus only the imaginary part contributes to the dynamical ``friction'' of interest. Substituting into Eq. (\ref{delta_rho_int}), the $p^2$'s cancel, making the integrand a total derivative. Therefore, only the boundary term at $p=m_\nu |\vc{v}_{\nu c}\cdot {\bf k}|/|{\bf k}|$ contributes to the integral:
\beq
\delta \rho_{\nu, {\bf k}} |_{\rm dyn. fric.} = - \frac{N_\nu m^4_\nu \Phi_{\bf k}}{2\pi \hbar^3}\frac{i \vc{v}_{\nu c}\cdot {\bf k} / |{\bf k}|}{\exp[(m_\nu c/T_\nu) (\vc{v}_{\nu c}\cdot {\bf k} / |{\bf k}|)]+1},  
\eeq
which using Gauss's law ${\bf \nabla \cdot g }= -4\pi G\rho$, leads to the gravitational field due to dynamical friction:
 \bea
 {\bf g}_{\nu, \bf k} |_{\rm dyn. fric.}  &=&  \frac{2N_\nu G m^4_\nu \Phi_{\bf k}}{  \hbar^3 |{\bf k}|^3}\frac{ (\vc{v}_{\nu c}\cdot {\bf k}) {\bf k}}{\exp[(m_\nu c/T_\nu) (|\vc{v}_{\nu c}\cdot {\bf k}| / |{\bf k}|)]+1} \nonumber \\
 &\simeq&   \frac{ 2  N_\nu G m^4_\nu \mu(|\vc{v}_{\nu c}|) \Phi_{\bf k} (\vc{v}_{\nu c}\cdot {\bf k}) {\bf k}}{  \hbar^3 |{\bf k}|^3}, \label{gnuc}
 \eea
where $ 0.7 \lesssim \mu  < 1$ captures the velocity dependence of the \textcolor{black}{exponential term in the denominator of first equation. We will assume $\mu= 1$ where necessary in our numerical evaluations.}

\begin{figure}
\begin{center}
\includegraphics[width=9cm]{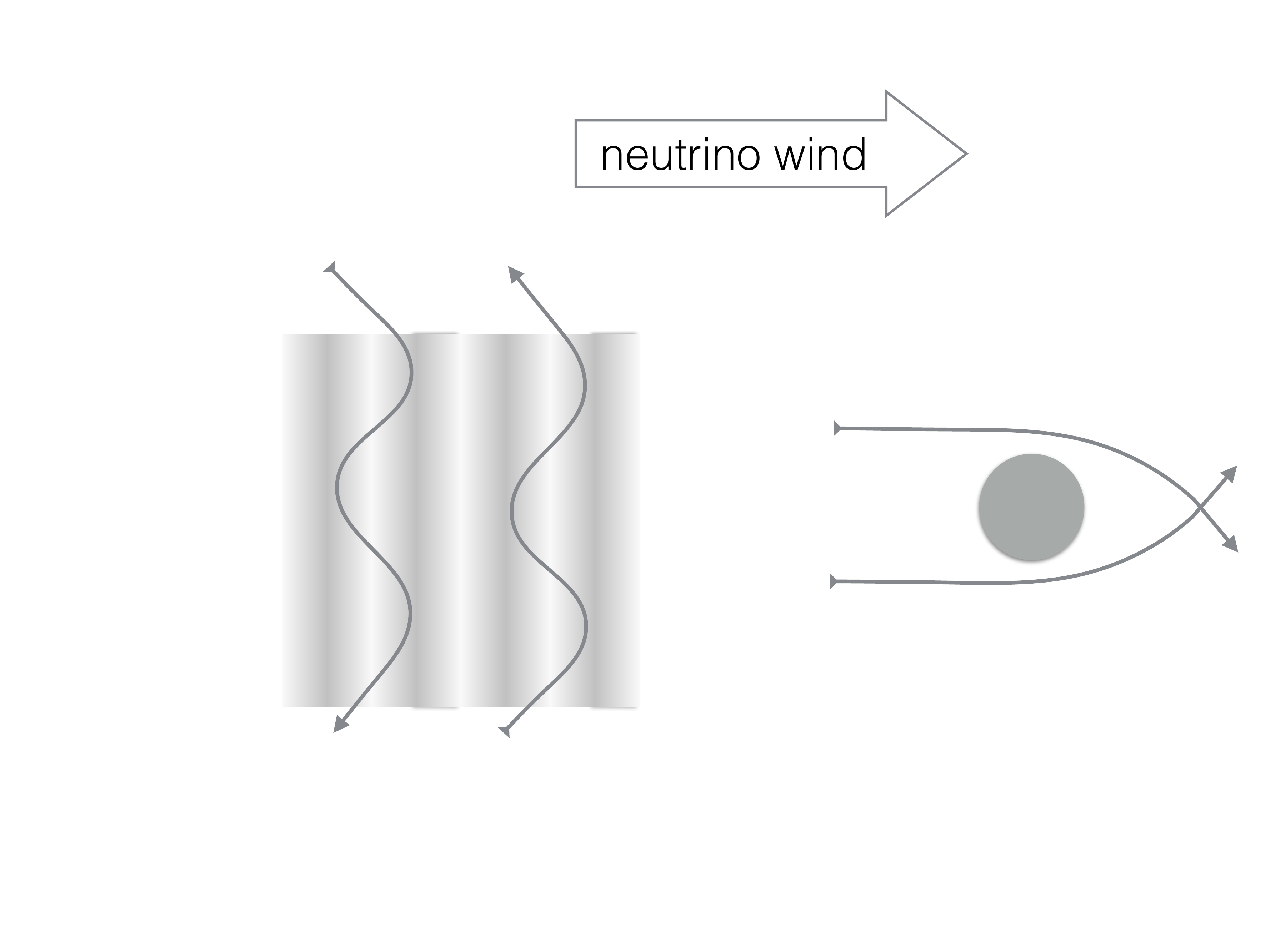}
\caption{A depiction of dynamical friction experienced by a plane wave packet (left) and  a compact halo (right), caused by neutrino wind (see text for details). } 
\label{plane_vs_sphere}
\end{center}
\end{figure}

Let us briefly discuss the physical picture of a neutrino wake for a plane wave. For a finite object such as a halo (or solid sphere discussed above), a neutrino wake would form downstream, due to gravitational focusing, which is responsible for slowing down the halo (hence the term, dynamical friction). The picture is qualitatively different for a CDM plane wave, as planar symmetry appears to prevent any focusing. The key here though is that, in nature, we always deal with finite wave packets, which do break exact planar symmetry. As the wave packet moves through the neutrino sea, the neutrinos with velocities nearly parallel to the wavefront (in the packet rest frame, corresponding to the singularity in Eq. (\ref{delta_rho_int})) can get trapped in between the potential peaks of the travelling wave. While the neutrinos that are trapped are preferentially travelling downstream, their velocity randomizes by the time they get out, implying that they impart a net momentum on the wave packet, in the direction of the wind (see Figure \ref{plane_vs_sphere}). Despite this qualitative difference, we shall see below that if we construct a spherical halo from the superposition of Fourier modes in Eq. (\ref{gnuc}), we shall recover the standard dynamical friction for an isolated sphere (Eq. (\ref{force_solid})).  

\subsection{Halo Model}
\label{section_halomodel}
The halo model provides a realistic description of the nonlinear CDM distribution, which includes modelling the profiles of individual haloes, as well as their clustering \textcolor{black}{\citep[see][for a general review of the halo model formalism]{cooray2002}}. As a result, clustered haloes will contribute to each other's wakes.
The dynamical friction force on a halo ${\bf F} |_{\rm dyn. fric.}$ is given by:
\bea
\vc{v}_{\nu c} &\cdot& {\bf F} |_{\rm dyn. fric.} = \int d^3 x~ \rho_{\rm halo}({\bf x})~  \vc{v}_{\nu c} \cdot  {\bf g}_{\nu}({\bf x}) |_{\rm dyn. fric.}  \nonumber \\ 
&=& \frac{\mu(|\vc{v}_{\nu c}|) N_\nu G m^4_\nu v_{\nu c}^i v_{\nu c}^j}{   \hbar^3} \int \frac{d^3 k}{(2\pi)^3} \Phi_k \rho_{{\rm halo}, k} \frac{k_i k_j}{k^3} \nonumber \\
&=&  \frac{4\pi \mu(|\vc{v}_{\nu c}|) N_\nu G^2 m^4_\nu v_{\nu c}^2}{ 3 \hbar^3} \int \frac{d^3 k}{(2\pi)^3} \frac{\rho_k \rho_{{\rm halo},k}}{k^3}, \\
\eea
where in the last two steps, we used spherical symmetry and Poisson's equation. Using the halo model, this can be written as:
\begin{align}
  \vc{v}_{\nu c} \cdot {\bf F} |_{\rm dyn. fric.} &=  \frac{4\pi \mu(|\vc{v}_{\nu c}|) N_\nu G^2 m^4_\nu v^2_{\nu c}}{ 3 \hbar^3} \int \frac{d^3 k}{(2\pi k)^3} \nonumber \\
&  \left[\rho^2_{{\rm halo},k}(M)+ \int d M' \frac{dn}{d M'} b(M) b(M')  \right. \nonumber \\
& \left. P_{\rm CDM}(k)  \rho_{{\rm halo},k}(M) \rho_{{\rm halo},k}(M') \frac{}{} \right],
\end{align}
\textcolor{black}{where $P_{\rm CDM}(k)$ is the {\rm CDM} linear power spectrum.} The formula for the acceleration due to dynamical friction \textcolor{black}{is then given as} 
\begin{align}
  {\bf a} |_{\rm dyn. fric.} &= M^{-1}_{\rm halo} {\bf F} |_{\rm dyn. fric.} \nonumber \\
  &=  \frac{2 \mu(|\vc{v}_{\nu c}|) N_\nu G^2 m^4_\nu  \vc{v}_{\nu c}}{ 3 \pi \hbar^3} \int \frac{d k}{k} \nonumber \\
  & \left[ \frac{}{}  M_{\rm halo} u(k|M_{\rm halo})^2+ b(M_{\rm halo}) P_{\rm CDM}(k) u(k|M_{\rm halo})  \right. \nonumber \\
  &  \int d M' \frac{dn}{d M'} b(M') M'  u(k|M') \left. \frac{}{} \right].
\end{align}
For the halo model, we use the convention of \cite{cooray2002} for the normalised Fourier transform of the halo profile (their eq. 107):
\beq
\label{u_k_M}
u({\bf k}|M) \equiv \frac{\int d^3{\bf x}~ \rho_{\rm halo}({\bf x}|M) \exp(-i {\bf k\cdot x}) }{\int d^3 {\bf x}~ \rho_{\rm halo}({\bf x}|M) },
\eeq
where $b(M)$ and $dn/dM$ are the linear bias and mass function, respectively, estimated using the fitting formula of  \cite{tinker2008,tinker2010} and the concentration-mass relation of \cite{okoli2016}. Our final predictions are not sensitive to the specifics of the structure formation \textcolor{black}{parameters such as the form of the density profile, the concentration-mass relation, the linear bias model, the mass function, etc.}
We will also assume the NFW halo mass profile proposed by \cite{navarro1996} that has the form:
\begin{equation}
\label{eq_NFW}
\rho_{\rm NFW}(r) = \frac{\rho_s}{(r/r_s)(1+r/r_s)^2},
\end{equation}
where $r_s$ is a characteristic scale radius (the radius at which the logarithmic slope of the density is $-2$ ), and $\rho_s$ is an inner density parameter. 

\begin{figure}
\begin{center}
\includegraphics[width=9cm]{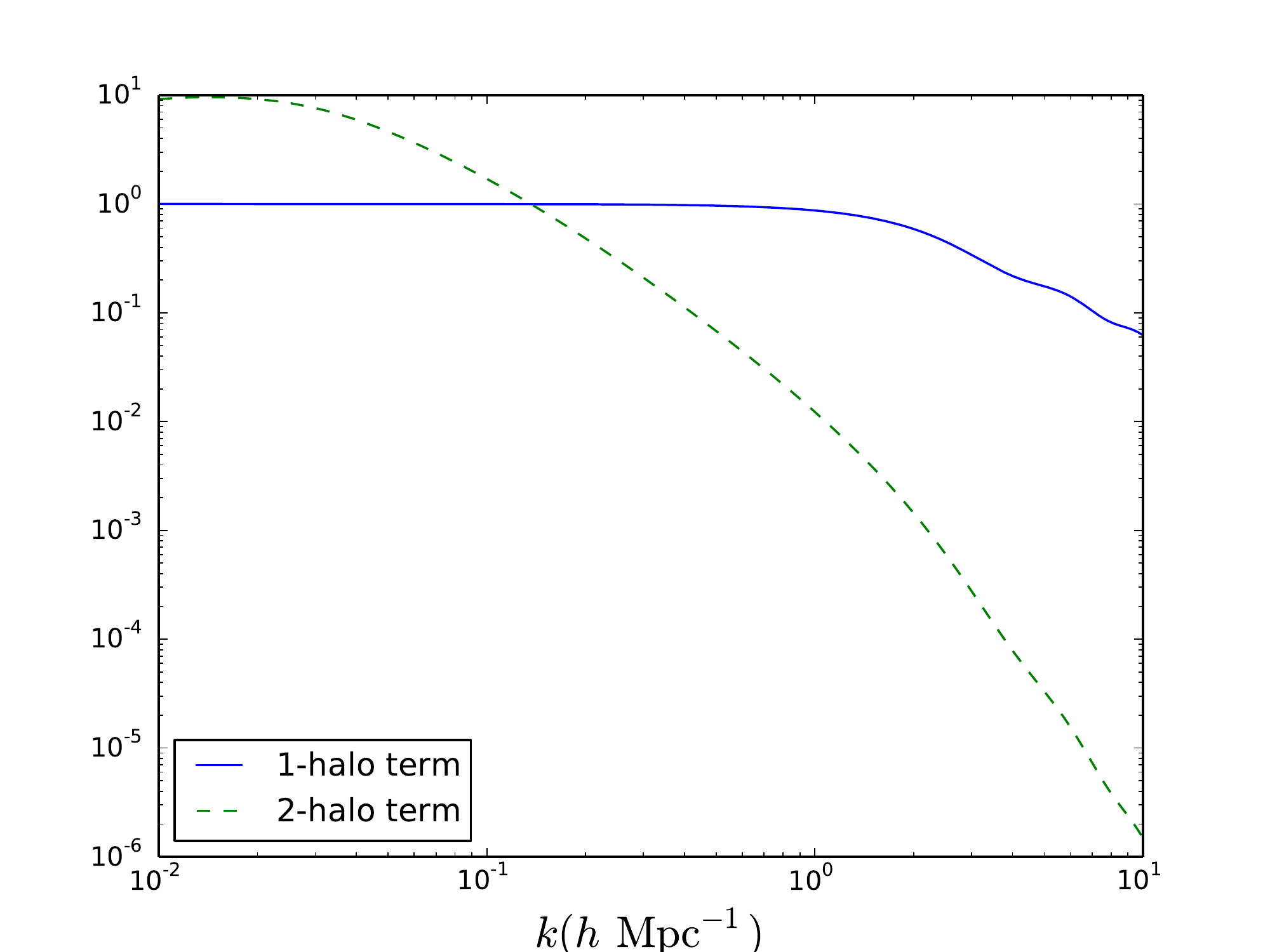}
\caption{The one-halo and two-halo terms in square brackets in Eq. (\ref{eq_deltav_halomodel}), as a function of $k$, for $M_{\rm halo} = 10^{15} h^{-1}\msol$ and a concentration of 4. } 
\label{um_integral}
\end{center}
\end{figure}

Incorporating the details of the halo model, the change in relative velocity for a dark matter halo in terms of the sum of neutrino mass is given by
\begin{align}
\label{eq_deltav_halomodel}
\frac{\Delta v_{\nu c}}{v_{\nu c}}  &= - \frac{2 \mu(\vc{v}_{\nu c},\vc{k}) G^2  M_{\rm halo} N_{\nu} m^4_{\nu}   }{ 3 \pi H \hbar^3 } \int \frac{d k}{k} \nonumber \\
  & \left[ \frac{}{} u(k|M_{\rm halo})^2+ \frac{b(M_{\rm halo}) P_{\rm CDM}(k) u(k|M_{\rm halo})}{M_{\rm halo}}  \right. \nonumber \\
  &  \int d M' \frac{dn}{d M'} b(M') M'  u(k|M') \left. \frac{}{} \right].
\end{align}
The first term in square brackets (the one-halo term) illustrates the contribution to dynamical friction by the main halo (identical to what we found for a solid sphere, Eq. (\ref{solid}), for $\mu=1$) while the second term (the two-halo term) describes that due to nearby haloes clustered with the main halo. We show the behaviour of the first and second parts of the term in square brackets in Figure ~\ref{um_integral}. 

\begin{figure}
\begin{center}
\includegraphics[width=9cm]{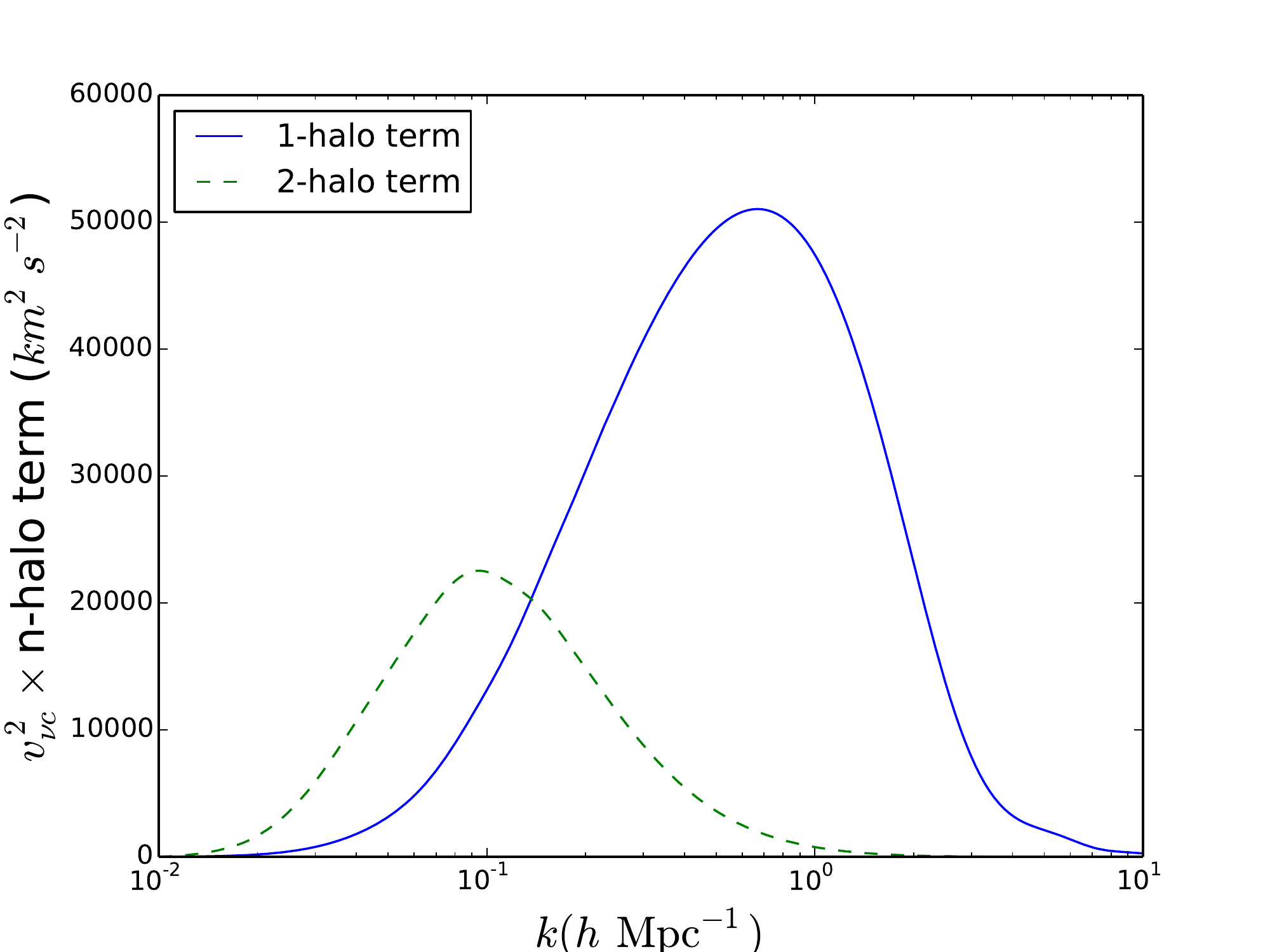}
\caption{The one-halo and two-halo terms of the term in square brackets including $[v'_{\nu c}(<k)]^2$ as in Eq. (\ref{v_nuc_inclv2}), for $M_{\rm halo} = 10^{15}h^{-1}\msol$ and a $0.2 \, \rm eV$ neutrino .} 
\label{m_integral}
\end{center}
\end{figure}

The integral over $k$ in Eq. (\ref{eq_deltav_halomodel}) is logarithmically divergent since both the one- and two-halo terms formally extend to infinity; suitable limits in $k$ must therefore be chosen. In order to set limits that are physically meaningful, we can multiply both sides of Eq. (\ref{eq_deltav_halomodel}) by $v_{\nu c}^2$ and incorporate this relative velocity inside the integral \textcolor{black}{(shown in Figure \ref{m_integral})}. At a given wavenumber $k$, the contribution to the neutrino-CDM relative velocity is only from larger length scales outside the the scale of the halo. We also introduce the sum of the different neutrino species. This form integrates the hierarchy of the neutrinos into our calculations and, relaxes the assumption of the degenerate neutrino species.
Thus, the equation for $\Delta v_{\nu c}$ becomes
\begin{align}
\label{v_nuc_inclv2}
\langle v_{\nu c} \Delta v_{\nu c} \rangle  &\simeq - \frac{ 2 \mu(\vc{v}_{\nu c},\vc{k}) G^2  M_{\rm halo}  }{ 3 \pi H \hbar^3 } \sum_{i=1}^3 m^4_i \int \frac{d k}{k} \langle [v'_{i c}(<k)]^2\rangle \nonumber \\
  & \left[ \frac{}{} u(k|M_{\rm halo})^2+ \frac{b(M_{\rm halo}) P_{\rm CDM}(k) u(k|M_{\rm halo})}{M_{\rm halo}}   \right. \nonumber  \\
  &  \int d M' \frac{dn}{d M'} b(M') M'  u(k|M')  \left. \frac{}{} \right],
\end{align}
where
\beq
\langle [v'_{i c}(<k)]^2 \rangle = a^2 \int_0^k \frac{dk'}{k'} \Delta^2_{v i c}(k') \widetilde{W}^2(k'R),
\eeq
and $i$ is for the different neutrino species and $v_{\nu c}$ is the relative velocity from all scales and from all neutrino species. 

\textcolor{black}{Recall that the relative velocity between the neutrino and dark matter decreases with scale and goes to zero on very large scales (Figure \ref{v_nuc_r}). On the other hand, the effects of nonlinearities are expected to be more significant on smaller scales. Therefore, for concreteness, we choose a mid-point of $16 h^{-1}{\rm Mpc}$ to filter relative neutrino-CDM velocity field $v_{\nu c}$. } 

\begin{figure}
\begin{center}
\includegraphics[width=9cm]{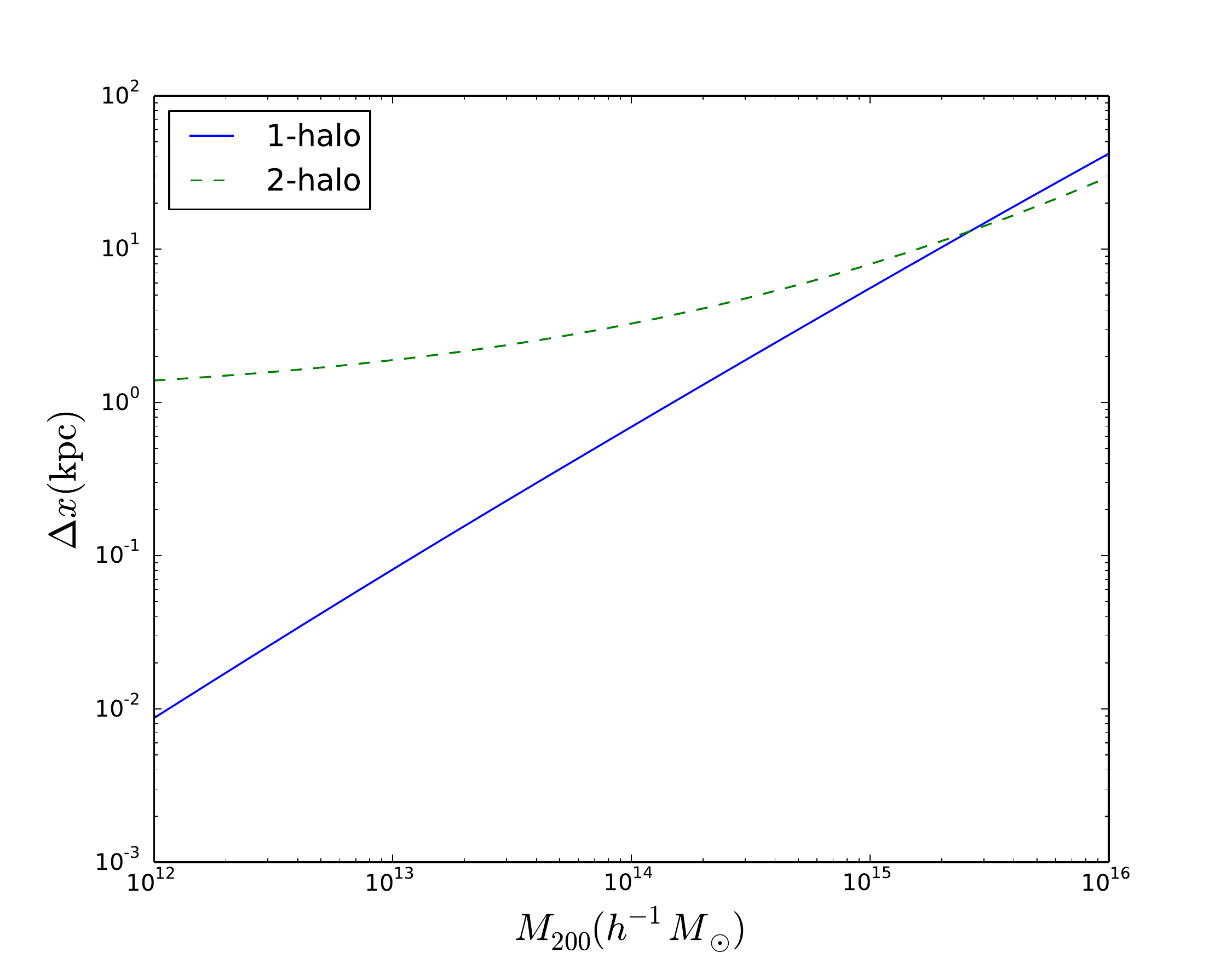}
\caption{The 1-halo and 2-halo contribution to the displacement due to dynamical friction for a $0.1 \, {\rm eV}$ neutrino. It is evident that the 2-halo term dominates for all masses less than $\sim 10^{15} h^{-1}\msol$ where the 1-halo term starts dominating.} 
\label{del_x_1_2}
\end{center}
\end{figure}

\begin{figure}
\begin{center}
\includegraphics[width=9cm]{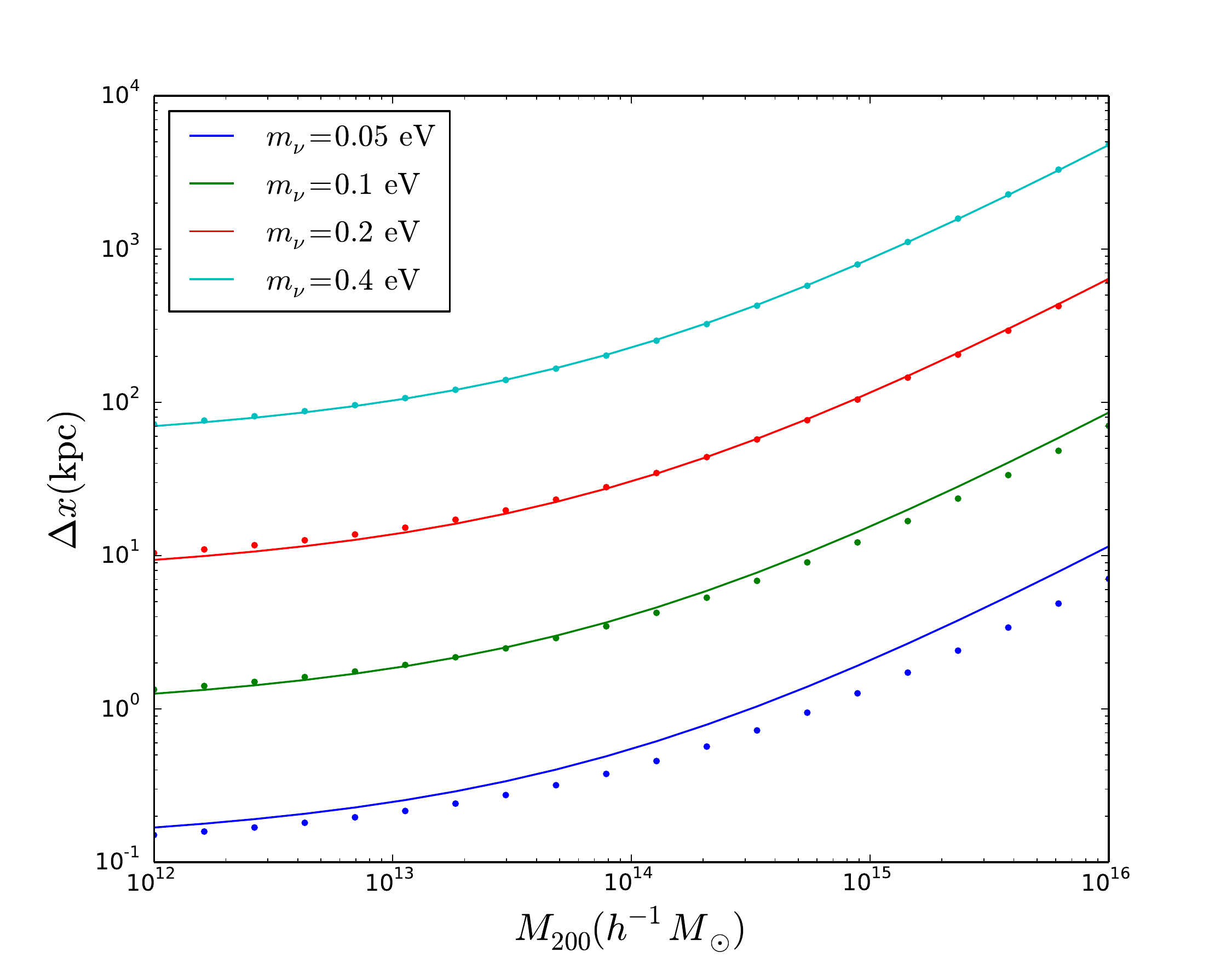}
\caption{A plot comparing the fitting function in Eq. (\ref{eq_fit}) with the calculated values for different neutrino masses. The dots are the calculated values while the line is the fitting function.} 
\label{fit_mh_mnu}
\end{center}
\end{figure}

%
%
A good fit (for a single neutrino) for the  drift velocity $\Delta v_{\nu c}$ and displacement $\Delta x \, (\simeq \frac{1}{2} \Delta v_{\nu c} \times t)$ \textcolor{black}{over a Hubble time} in terms of the neutrino mass $m_{\nu}$ and the halo mass $M_h$ is given by 
\begin{eqnarray}
\Delta v_{\nu c} &\simeq& \frac{\langle v_{\nu c} \Delta v_{\nu c} \rangle}{\langle v_{\nu c}^2 \rangle}  v_{\nu c}\\ &\simeq& (0.2 ~{\rm km/s})  \left[b(M_h) + \left(\frac{M_h}{1.3 \times 10^{14}h^{-1}M_{\odot}}\right)^{0.85} \right]\nonumber  \\
&& \left(\frac{m_{\nu}}{0.1 ~{\rm eV}}\right)^{2.9}  \left( v_{\nu c, 16} \over 193 ~{\rm kms^{-1}}\right),\\
\Delta x &\simeq& (1.5~{\rm kpc})  \left[b(M_h) + \left(\frac{M_h}{1.3 \times 10^{14}h^{-1}M_{\odot}}\right)^{0.85} \right] \nonumber\\
&& \left(\frac{m_{\nu}}{0.1 ~{\rm eV}}\right)^{2.9}\left( v_{\nu c, 16} \over 193 ~{\rm kms^{-1}}\right), 
\label{eq_fit}
\end{eqnarray}
where $v_{\nu c, 16}$ is the relative velocity of neutrinos and CDM, averaged over a sphere of radius $16 h^{-1}{\rm Mpc}$, centered around the halo. The explicit dependence of the displacement, $\Delta x$ on the 1-halo and 2-halo terms as a function of mass is shown in Figure \ref{del_x_1_2}. \textcolor{black}{Figure \ref{fit_mh_mnu} shows a comparison between $\Delta x$ and the best fit in Eq. (\ref{eq_fit}).}

\section{Predicted signal-to-noise for nominal surveys}
\label{sec:sigtonoise}

\begin{figure}
\begin{center}
\includegraphics[width=10cm]{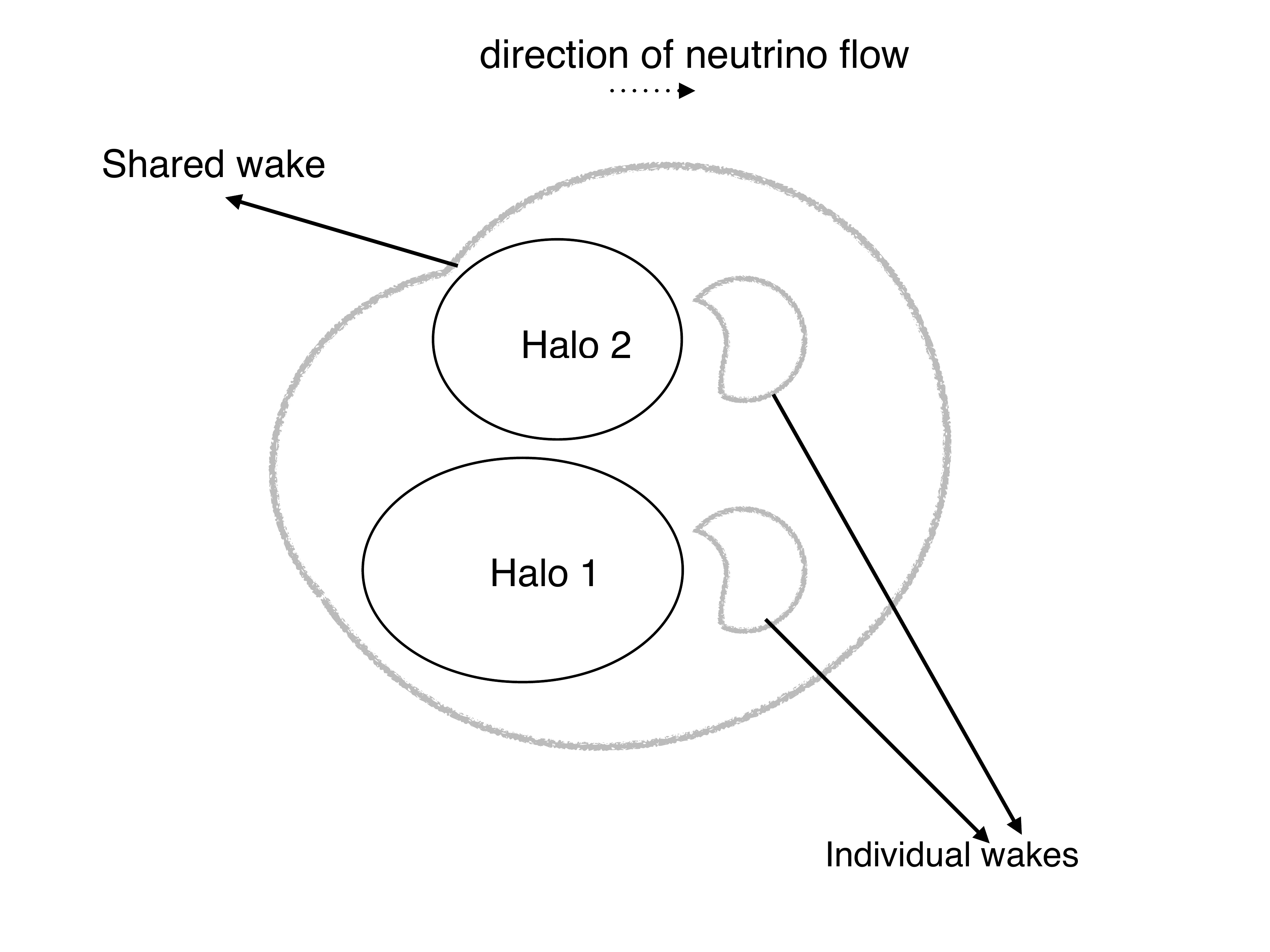}
\caption{A sketch of the effect of shared wakes on the displacement of two haloes.} 
\label{wake_cartoon}
\end{center}
\end{figure}

In this section, we predict the observational prospects for the detection of neutrino dynamical friction in galaxy surveys. One may be tempted to interpret Eq. (\ref{eq_fit}) as a relative displacement/velocity of haloes of different mass due the drag by the neutrino wind. However, this is only correct given the assumption that the haloes are not in the same neighbourhood and thus have their individual wakes. To make the theoretical predictions in Eq. (\ref{eq_fit}), we have averaged over all the haloes distributed around a given halo. In practice, this would lead to an insignificant signal in cross correlating two haloes in the same neighbourhood. The reason for this infinitesimal signal is that they share the same large scale neutrino wake (depicted in Figure \ref{wake_cartoon}), and thus only experience a small fraction of $\Delta v_{\nu c}$. To achieve an acceptable signal-to-noise estimate, we shall focus on the effect on the large scale distribution of haloes due the gravity of the neutrino wake. 

Recall that the effect of dynamical friction on the gravitational field of the halo is given by Eq. (\ref{gnuc}), which is repeated here for convenience, 
 \bea
 {\bf g}_{\nu, \bf k} |_{\rm dyn. fric.}  &\simeq&   \frac{ 2   G m^4_\nu \mu(|\vc{v}_{\nu c}|) \Phi_{\bf k} (\vc{v}_{\nu c}\cdot {\bf k}) {\bf k}}{  \hbar^3 |{\bf k}|^3}\nonumber.
 \eea
 
This extra effect modifies the total density of gravitating matter in Fourier space, including the neutrino hierarchy, as 
\begin{eqnarray}
\delta_{m,k} &\rightarrow& \delta_{m,k}  \left(1+i \frac{2a^2G  \sum_{i=1}^3 m^4_i  \mu {\bf v}_{i c}(z) \cdot {\bf k}}{\hbar^3 |\bf{k}|^3}\right), \nonumber \\
 &=& \delta_{m,k}  \left(1+i \phi_{\bf k}\right),
\label{eq:phase}
\end{eqnarray}
where 
\begin{equation}
\phi_{\bf k} \equiv \frac{2a^2 G  \sum_{i=1}^3 m^4_i  \mu {\bf v}_{i c}(z) \cdot {\bf k}}{\hbar^3 |\bf{k}|^3}
\label{eq:phasedef}
\end{equation}

Thus, the density is modified by an extra time-dependent phase that manifests through the time dependence of the $\nu$-CDM relative velocity. We aim to measure this consequence of dynamical friction due to neutrinos. To this end, we cross-correlate the densities of two tracers (possibly galaxies with different biases) in redshift space. The signal appears as an imaginary term in the redshift-space cross correlation spectrum (see Appendix~\ref{appendix_SN} for more details). The signal-to-noise for such a measurement is proportional to the difference in bias for the two populations, $(b_l - b_f)$, the effective volume of the survey $V_{\rm eff}$ and the time derivative of the phase term in Eq. (\ref{eq:phasedef}).

Let us make some simple theoretical predictions for signal-to-noise for a generic redshift survey. For a given survey with two tracers that have differing bias such as ``luminous galaxies'' (l) and ``faint galaxies'' (f), 
the signal to noise for the imaginary part of the galaxy-cluster redshift-space power spectrum due to the neutrino dynamical friction of the haloes can be calculated to be (see Appendix~\ref{appendix_SN} for further details):
\beq
\left(\frac{S}{N}\right)_{RSD}^2 = \frac{2  V_{\rm eff}}{(2\pi)^2} \int^{k_{\rm max}} d k k^2 P_k^2 \Delta b^2 \frac{\langle |\dot{\phi}_{\bf k}^2| \rangle}{5H^2} |\det C|^{-1} ,
\eeq
where $\Delta b$ is the relative linear bias between the two tracers \textcolor{black}{(we include only the linear bias and consider the nonlinear bias a part of the nonlinearities in structure formation under Section \ref{sec:discussion})}, $P_k$ is the power spectrum of matter fluctuations and $V_{\rm eff}$ is the effective survey volume\footnote{For the integral, we have assumed $k_{\rm max} =1 h^{-1}$Mpc.}. We see that the S/N increases with the square root of the survey volume and is larger for tracers with considerable relative bias. 
\begin{figure}
\begin{center}
\includegraphics[width=9cm]{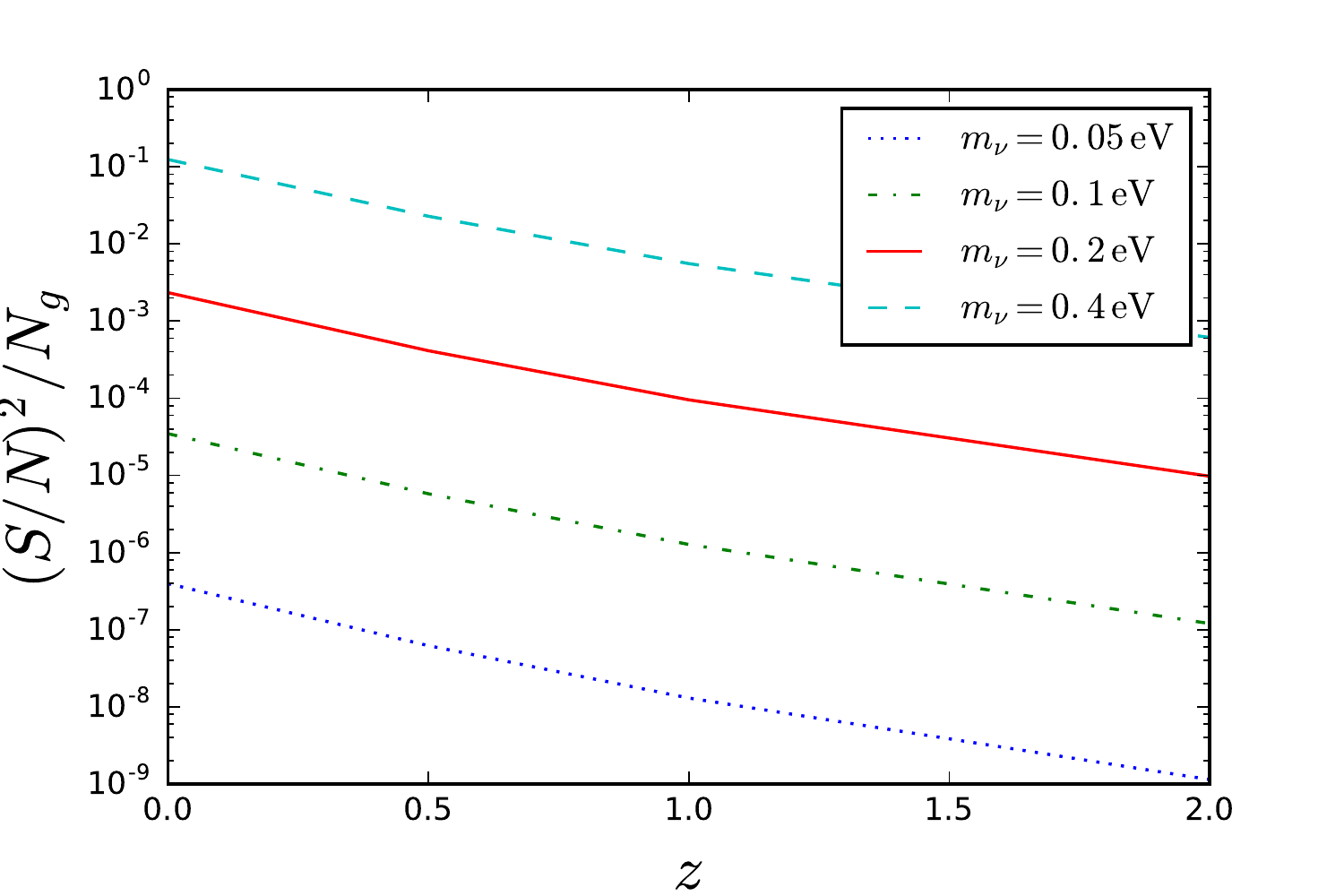}
\caption{The signal-to-noise squared per galaxy as a function of redshift for various neutrino masses. This signal is estimated with a number density $n_l \sim n_f = 0.02 \, h^3 \rm Mpc^{-3}$.}
\label{SN_sq}
\end{center}
\end{figure}

Figure \ref{SN_sq} shows our projection for $(S/N)^2$ per galaxy for the detection of the imaginary part in the redshift-space cross-power spectrum of two tracers (in a nominal survey) with a relative bias $\Delta b \sim 1$ \textcolor{black}{(a typical relative bias between luminous and faint galaxies e.g \citet{skibba2014})}, as a function of redshift. These estimates are made over a filtering comoving radius of $16~h^{-1}{\rm Mpc}$. \textcolor{black}{The $(S/N)^2$ per galaxy} decreases as the redshift increases due to the steep drop in $\dot{\phi}_{\rm k}$. For a single neutrino of mass $0.07 \, \rm eV$,  a survey with about 2 million galaxies at $z \lesssim 0.5$ can achieve a 2-3$\sigma$ detection. This may be already achievable by the SDSS main sample \citep{2009ApJS..182..543A} or BOSS redshift surveys \citep{2014ApJS..211...17A}, and will be improved by a factor of $3$, in the upcoming DESI survey \citep{2013arXiv1308.0847L}. 
Signals from higher neutrino masses due to dynamical friction effects from the neutrino distribution are significantly easier to extract, with $(S/N)^2 \propto m_\nu^6$. For a generic redshift survey, we project that
\beq
N_g \gtrsim 1.7 \times 10^7 \left(m_\nu \over 0.05 ~{\rm eV}\right)^{-6}\frac{28.5^z}{(N_\nu\Delta b)^2} \label{n_gal}
\eeq
galaxies can lead to  $> 3\sigma$ detection of the neutrino drag.

\begin{figure}
\begin{center}
\includegraphics[width=9cm]{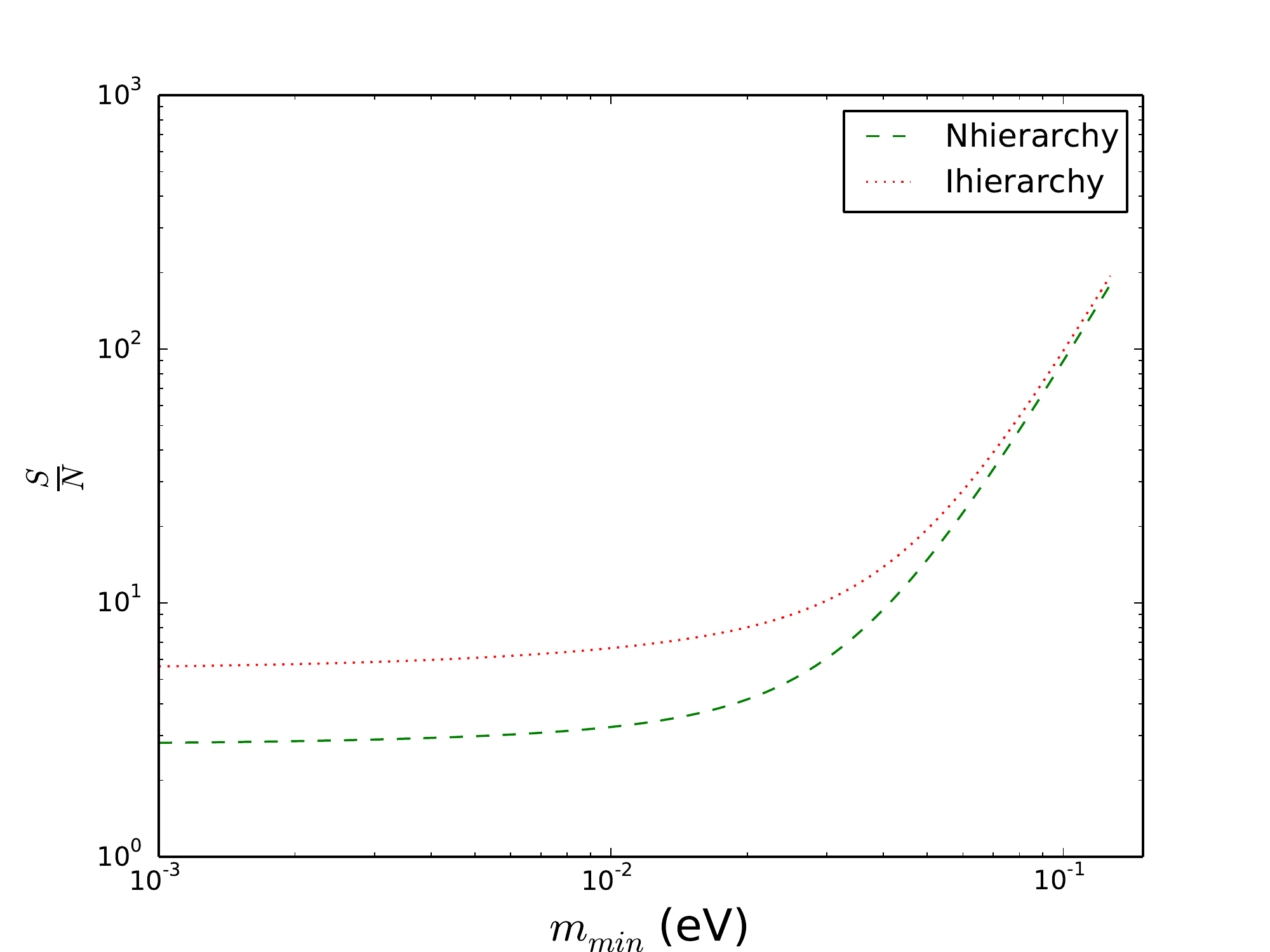}
\caption{S/N for the detection of the imaginary part of the redshift-space cross power spectrum of two tracers due to the dynamical friction effect of massive neutrinos. The dashed and dotted lines represents the S/N for the sum of the neutrino masses assuming the normal and inverted hierarchies respectively. Both plots are for a theoretical survey assuming a volume of $V_{\rm eff}=1 h^{-3} {\rm Gpc}^3$, $\Delta b = 1$ and $n_l \sim  n_f = 0.02~ h^3 \rm Mpc^{-3}$. } 
\label{SN}
\end{center}
\end{figure}

\begin{figure}
\begin{center}
\includegraphics[width=9cm]{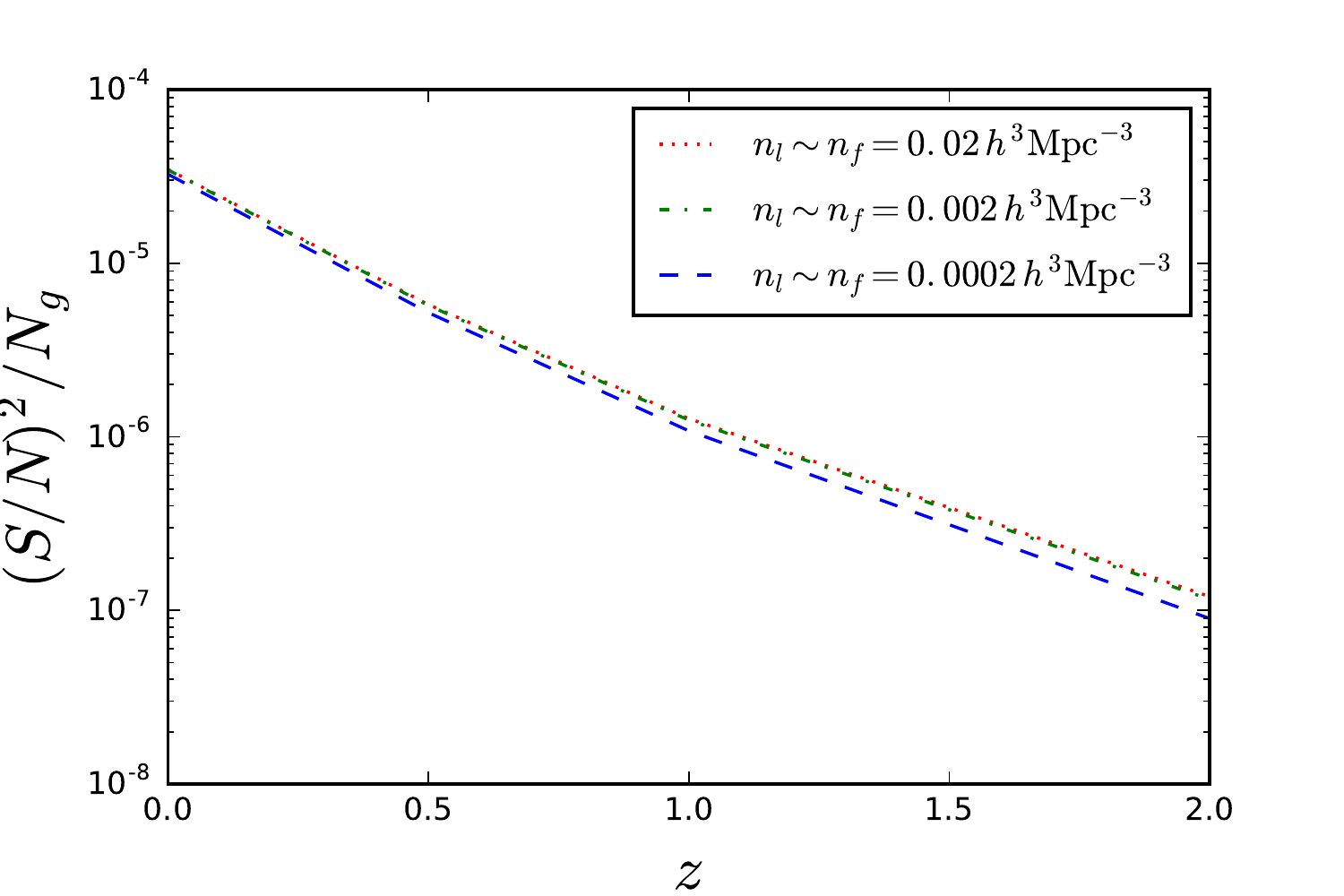}
\caption{The signal-to-noise squared per galaxy as a function of redshift for various number densities, $n_l$ and $n_f$, \textcolor{black}{expected from the SPHEREX all sky survey \citep{2014arXiv1412.4872D}}. This signal is estimated for a $0.1 \, \rm eV$ neutrino.}
\label{SNsq}
\end{center}
\end{figure}

For constraints on the sum of neutrino masses, $M_{\nu}$, one will have to sum over the signal from the various contributing species. Figure \ref{SN} shows these estimates as a function of the minimum neutrino mass (which can be easily related to the sum of neutrino masses, $M_{\nu}$) for the normal and inverted neutrino hierarchies. \textcolor{black}{Even though we have assumed the number density of the luminous and faint galaxies to be $\sim 0.02 \, h^3 \rm Mpc^{-3}$, Figure \ref{SNsq} confirms that the signal-to-noise squared per galaxy is \textcolor{black}{insensitive} of the number density of galaxies. Thus, the total signal-to-noise squared is proportional to the number of galaxies $N_g$,  $\Delta b^2,$ and $m_{\nu}^6$.}

An alternative measurement of the dynamical friction effect from neutrinos comes from a direct measurement of the velocities of galaxies, e.g., using the kinematic Sunyaev-Zel'dovich (kSZ) effect \citep[e.g.,][]{2013MNRAS.430.1617L,2016PhRvD..93h2002S,2016arXiv160702139D}. In this case, the $(S/N)^2$ is given as 
\begin{equation}
\left(\frac{S}{N}\right)_{\rm kSZ}^2 = \sum_{N_g} \frac{\Delta v_{\nu c}^2}{\sigma_v^2},
\end{equation}
with $\Delta v_{\nu c}$ given by Eq. (\ref{eq_fit}) and $\sigma_{v} \sim 1500 \, \rm km/s.$ is the velocity error per galaxy, expected from the future CMB S4 kSZ measurements \citep[e.g.,][]{2016arXiv160606367S}. For a $0.1\, \rm eV$ neutrino and a halo of mass $M_h  \sim 10^{13} h^{-1} M_{\odot}.$, in future  spectroscopic surveys such as DESI, EUCLID and SPHEREX (spectro), with $\lesssim 10^8$ galaxies, we only get $S/N \lesssim 1$.  Therefore, the next generation of kSZ surveys will not be an efficient probe of neutrino dynamical friction. 

In Appendix \ref{sec:nonlinearities}, we look at the contamination of the relative velocity signal by nonlinear structure formation using $\Lambda$CDM N-body simulations, which can become comparable to kSZ measurement errors. \textcolor{black}{We find that the detection of the dynamical friction effects from neutrino require about $0.1\%$ precision in the measurement of the bulk velocity. Although current experiments do not have this precision, the required precision may be met with future experiments. }

\section{Discussion and Conclusion}
\label{sec:discussion}
We have investigated the prospects for detecting a novel signal from the cross-correlation of different galaxy populations in redshift space expected in the presence of neutrinos. This is due to the dynamical friction drag experienced by dark matter haloes that move in the primordial neutrino sea. Even though the neutrinos and dark matter cannot be observed, these effects make imprints on the galaxies and should be detectable in future surveys. 
With current surveys, a high \textcolor{black}{$(\gtrsim 10)$} S/N is predicted for  $M_{\nu} \sim 0.2\,\rm{eV}$, which are marginally allowed by a combination of galaxy surveys and CMB. Given the current limits on the neutrino mass, future generations of  high-density redshift surveys such as DESI  will be able to detect smaller mass neutrinos or sum of masses $M_{\nu} < 0.1\,\rm{eV}$. 

We should note that gravitational redshift can also introduce an imaginary part to the redshift space cross-power spectrum \citep{Mcdonald2009}. However, the signal from this effect is smaller than that from the effect of dynamical friction by neutrinos (for a $0.1 \, {\rm eV}$ neutrino) and has a different scale dependence ($\propto k^{-1}$ vs the scale-dependent neutrino signal that peaks at $k\sim 0.01$ Mpc/$h$). 

Other effects of neutrinos on large-scale structure have been discussed by \cite{jimenez2016}. Their study serves as a method for distinguishing the mass splitting of the neutrinos given a measured constraint on the sum of neutrino masses $M_{\nu}.$ The authors consider the effect of different neutrino masses on the power spectrum on different scales -- power suppression on small scales and the change of the matter-radiation equality scale on large scales. They also claim that this method may be used to distinguish the mass splitting of neutrinos given a precise measurement of the matter power spectrum which constrains the sum of neutrino mass. These effects become measurable on large linear scales and do not suffer from nonlinearities and systematic effects. Thus, given a measured constraint on the sum of neutrino masses, this method gives another independent confirmation/test of the mass splitting seen in neutrino oscillation experiments. Yet another impact of neutrinos on large-scale structure was studied by \cite{loverde2016}, who proposes a method in which neutrino mass may be constrained by the measurement of the scale-dependent bias and the linear growth parameter in upcoming large surveys. Notably, similar to our proposal, this measurement is not limited by cosmic variance. 

We next discuss some of the major systematics which may affect our predictions for the signal.

\subsection{Galaxy and bias}
Our estimates for the S/N are based on the number density of galaxies in a survey. Using galaxies requires a good estimate of the galaxy bias. Details and precision in defining the galaxy bias \citep[with respect to the total matter in the presence of neutrinos or to the cold dark matter alone;][]{castorina2014} are required in constraining the mass of neutrinos through the suppression of power on small scales. \textcolor{black}{ However, we expect the exact definition of bias to be less important in extracting the dynamical friction effect, given the large scale coherence of neutrino wind.}  

The extra scale dependence of the halo bias which may be due to nonlinearities and the free streaming scale of the neutrino is an extra observable proposed by \cite{loverde2016}. However, these affect the magnitude, not the phase, of the Fourier amplitudes at percent level, and thus should not bias our projections for neutrino drag. 

\subsection{Nonlinearities \textcolor{black}{in structure formation}}
The impact of nonlinearities on the CDM-neutrino relative velocity was investigated by \cite{inman2015} in \textcolor{black}{a number of $\nu \Lambda$CDM  simulations}. The authors were interested in the effectiveness of the simple linear theory approximation given the nonlinear complexities of structure formation. Their measurements \textcolor{black}{show} that the relative velocity power spectra predicted from linear theory are higher than that in simulations\textcolor{black}{, but are still within 30\% of each other for the empirically allowed neutrino masses}. Reconstructing the relative velocity power spectra using the halo density field and the dark matter density field, the authors show that they are correlated with the simulations for scales $k \lesssim 1h/\rm{Mpc}$ and also have the right direction of the relative velocity with a mass-dependent  correlation coefficient. The magnitude of the reconstruction may be corrected for nonlinearities by the ratio of the nonlinear to linear CDM power spectra. This reconstruction procedure may be implemented in practice to estimate the full neutrino-CDM relative velocity power spectrum.

A more serious issue is whether nonlinear effects in standard nonlinear structure formation can mimic the effect of dynamical friction by neutrinos. After all, the modulation of galaxy (cross-)power spectrum, or relative velocity, by reconstructed $v_{\nu c}$ can be interpreted as a particular contribution to the galaxy bispectrum, which might be partially degenerate with the (much larger) nonlinear halo bispectrum. While Appendix \ref{sec:nonlinearities} provides a first look at the magnitude of this degeneracy, a more complete study of this degeneracy in $\nu \Lambda$CDM simulations \citep{inman2015} is currently underway. \textcolor{black}{Indeed, our preliminary analysis suggests that nonlinear effects are not degenerate with, and only have a marginal impact on our predictions.}

\subsection{Prospects for detection}
In practice, measuring the mass of neutrinos from the dynamical friction effect will be quite challenging, and requires a good knowledge and control of standard nonlinear structure formation, neutrino effects, and halo bias. Our $S/N$ estimates have only included the statistical error, and control of systematic uncertainties will only come from a careful study of simulated haloes. Nevertheless, our statistical projection of $S/N \gtrsim 3$ for future surveys (see Eq. \ref{n_gal} or Figure \ref{SN_sq}) provides an incentive for further theoretical study and improvement. 
The dynamical friction effect, described here, may be a powerful complement to the various ways in which neutrinos will be probed over the next decade.

\section*{Acknowledgements}
We thank Derek Inman, Dustin Lang, Marilena LoVerde, Ue-Li Pen, and Rafael Sorkin for valuable discussions. \textcolor{black}{We also thank the anonymous referee for comments that improved the presentation of this work.} This work is supported by the University of Waterloo and the Perimeter Institute for Theoretical Physics. MJH acknowledges support from NSERC. Research at the Perimeter Institute is supported by in part by the Government of Canada through Innovation, Science and Economic Development Canada and by the Province of Ontario through the Ministry of Research, Innovation and Science.

\appendix

\section{Signal-to-noise from the gravitational field due to dynamical friction}
\label{appendix_SN}
Following the correction to the the gravitational field due the dynamical friction effect from neutrinos, ${\bf g_k} \rightarrow {\bf g_k} + {\bf g}_{\nu, {\bf k}}$, we can express this as a measurable effect to the power spectrum of galaxies in a redshift survey. Using divergence form of Gauss's equation for gravity, $\nabla \cdot {\bf g} = - 4\pi a G  \rho$, the Fourier space version is given as 
\begin{equation}
{\bf g_k} = \frac{4 \pi i a G \rho_k {\bf k}}{\bf{|k|^2}}, 
\end{equation}
while that of the Poisson equation is 
\begin{equation}
\Phi_k = -\frac{4 \pi a^2 G \rho_k}{|\bf{k}|^2}.
\end{equation}
Thus, 
\begin{eqnarray}
{\bf g_k} &\rightarrow& {\bf g_k} + {\bf g}_{\nu, {\bf k}} \nonumber \\
 &=& \frac{4 \pi i a G \rho_k \bf{k}}{\bf{|k|^2}} \left(1+i \frac{2a^2G  \sum_{i=1}^3 m^4_i \mu {\bf v}_{i c}(z) \cdot {\bf k}}{\hbar^3 |\bf{k}|^3}\right),  \nonumber \\
&=& \frac{4 \pi i a G \rho_m \delta_k \bf{k}}{\bf{|k|^2}} \left(1+i \frac{2a^2 G\sum_{i=1}^3 m^4_i \mu {\bf v}_{i c}(z) \cdot {\bf k}}{\hbar^3 |\bf{k}|^3}\right), \nonumber \\ \label{eq:new_gravity}
\end{eqnarray}
where we have used Eq. (\ref{gnuc}) in comoving coordinates to substitute for $ {\bf g}_{\nu, {\bf k}}$. 

We shall see that this signal can be extracted by looking at the redshift space distortion of galaxy surveys. Recall that the redshift-space distance $r^s$ is related to the real-space distance $r^d$ through
\begin{equation}
r^s  = r^d + \frac{v_x}{aH},
\end{equation}
where $v_x$ is the velocity along the line of sight and $H$ is the Hubble constant.  Using continuity equation for matter, we can find the redshift space overdensity for galaxies $\delta^s_g$ in terms of its real space overdensity $\delta^d_g$ and matter overdensity $\delta_m$:
\beq
\delta_{g,k}^s = \delta_{g,k}^d - \nabla \cdot \left(\frac{v_x \hat{{\bf x}}}{aH}\right) \Rightarrow \delta_{g,k}^s = \delta_{g,k}^d - \frac{ { k_x^2} \dot{\delta}_{m,k}}{H |{\bf k}|^2 }.
\label{eq:rsd}
\eeq
Due to the dynamical friction of the neutrinos (assuming a uniform ${\bf v}_{\nu c}$) the matter overdensity has a redshift-dependent phase (\ref{eq:new_gravity}):
\begin{eqnarray}
\delta_{m,k}(z) &\simeq & D_L(k,z) e^{i \phi_{\bf k}(z)} \delta_{m0,k},\nonumber \\
\phi_{\bf k}(z) &=& \frac{2a^2 G  \sum_{i=1}^3 m^4_i \mu {\bf v}_{i c}(z) \cdot {\bf k}}{\hbar^3 |\bf{k}|^3},  \label{phase_shift}
\end{eqnarray}
where $D_L(k,z)$ is the (scale-dependent) linear growth factor for $\nu\Lambda$CDM cosmology. For a single tracer in redshift-space, the density perturbation is given as
\begin{eqnarray}
\delta_{g,k}^s &=& \delta_{g,k}^d - \frac{ k_x^2 \dot{\delta}_{m,k}}{H |{\bf k}|^2 } \nonumber \\
&=& \delta_{g,k}^d \left(1 - \frac{k_x^2}{b_g H |{\bf k}|^2 } \frac{\dot{D_L}}{D_L} - i \frac{\dot{\phi} k_x^2}{H b_g |{\bf k}|^2}\right)\nonumber \\
&=& \delta_{m,k}\left(b_g - \frac{k_x^2}{H |{\bf k}|^2 } \frac{\dot{D_L}}{D_L} - i \frac{\dot{\phi} k_x^2}{H|{\bf k}|^2}\right).
\end{eqnarray}
The first two terms in the above equation are the standard RSD terms while the last term is new, as a result of the effect of dynamical friction due to neutrinos. Measuring this signal from a single tracer requires a very good knowledge of the linear bias term and is also susceptible to sample variance. One may eliminate the sample variance limitation by considering multiple tracers, as first suggested in \citet{macdonald2009}. This is because, in the absence of shot noise, the ratio of Fourier amplitudes for two tracers can be measured perfectly, even for a single mode. While this ratio is real in standard multi-tracer RSD \citep{macdonald2009}, it finds an imaginary part (leading to a dipole) in the presence of a neutrino wind.  There could also be a comparable imaginary term for different tracer biases, $b_f$ and $b_l$, which we shall ignore for our simple S/N estimate here.

We will consider cross-correlating two distinct populations of galaxies with similar comoving densities $n_l$ and $n_f$, and biases $b_l$ and $b_f$. This is given as 
\begin{eqnarray}
C_{fl} &\equiv& \langle \delta_{l,k}^s \delta_{f,k}^{*s}\rangle = \left\langle \left(\delta_{l,k}^d - \frac{ { k_x^2} \dot{\delta}_{m,k}}{H |{\bf k}|^2 }\right)\left(\delta_{f,k}^{d *} - \frac{a{ k_x^2} \dot{\delta}_{m,k}^ *}{H |{\bf k}|^2 }\right) \right\rangle, \nonumber \\
&=& b_l b_f P_k\left[ 1 - \frac{\dot{D_L} { k_x^2}}{H D_L |{\bf k}|^2}\left(\frac{1}{b_l} + \frac{1}{b_f}\right)  \right. +  \nonumber \\
&& \frac{ { k_x^4}}{b_l b_f H |{\bf k}|^4} \left(\frac{\dot{D_L^2}}{D_L^2} + \dot{\phi}_{\bf k}^2\right)\left. \frac{}{} \right] - \frac{i b_l b_f P_k\dot{\phi}_{\bf k} { k_x^2} }{H |{\bf k}|^2} \left(\frac{1}{b_l} - \frac{1}{b_f}\right).\nonumber \\
\end{eqnarray}
As can be seen from the last equation, the dynamical friction due to neutrinos has introduced an imaginary term to the redshift- space cross-power spectrum. 
This imaginary term is the signal per mode we hope to capture. Recall that 
\beq
\dot{\phi}_{\bf k} = \sum_{i=1}^3 \left(\frac{2G m^4_i \mu H }{\hbar^3 |\bf{k}|^3}\right)  {\bf k} \cdot \left[2 a^2{\bf v}_{i c}(z)-a\frac{\partial {\bf v}_{i c}(z)}{\partial z}\right].
\eeq

Ignoring the effect of neutrinos, the noise variance for every mode is given by \citep[e.g.,][]{Mcdonald2009}:
\bea
 (\Delta {\rm ~Im~} C_{fl})^2 &=&  \langle |{\rm ~Im~} C_{fl} |^2\rangle = \frac{1}{2}(C_{ff}C_{ll} - C_{fl}C_{fl}) \nonumber \\
 C_{ll} &=& \left(b_l -   \frac{{ k_x^2} \dot{D_L}}{ |{\bf k}|^2 H D_L } \right)^2 P_k +n^{-1}_l \nonumber \\
 C_{ff} &=& \left(b_f -   \frac{{ k_x^2} \dot{D_L}}{ |{\bf k}|^2 H D_L } \right)^2 P_k +n^{-1}_f \nonumber \\
 C_{fl} &=& \left(b_l -   \frac{{ k_x^2} \dot{D_L}}{ |{\bf k}|^2 H D_L } \right)\left(b_f -   \frac{{ k_x^2} \dot{D_L}}{ |{\bf k}|^2 H D_L } \right)P_k\nonumber \\
\eea
Consider these as the elements of a matrix $C$ given by
\[
\begin{bmatrix}
C_{ll}  &  C_{fl} \\
 C_{fl} &  C_{ff}
\end{bmatrix}
\] the noise variance is thus one half of the determinant of C, $\det C$.

\begin{figure}
\begin{center}
\includegraphics[width=9cm]{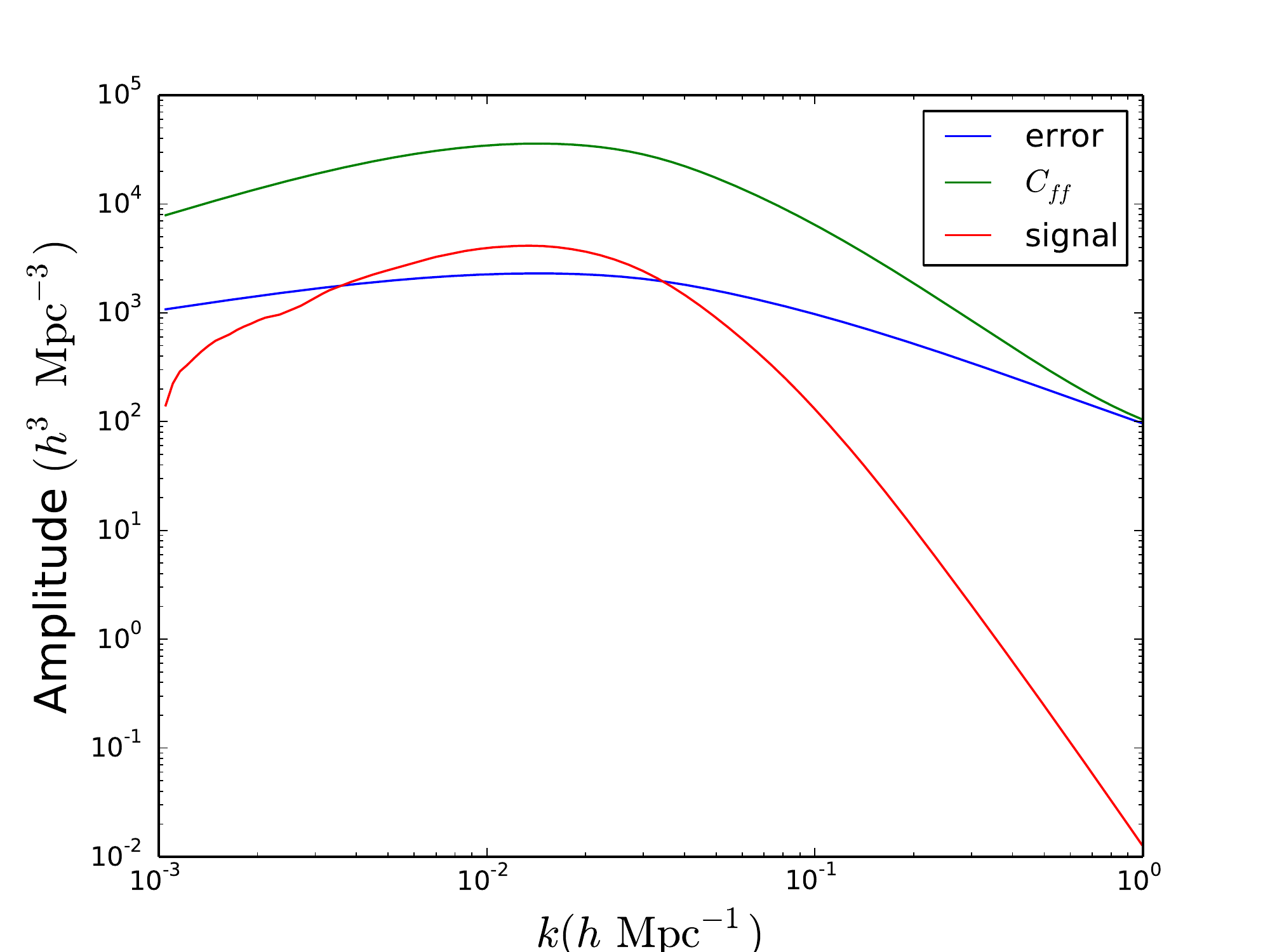}
\caption{The amplitude of the signal from the imaginary part of the cross-power spectrum, the error on the signal, and the auto-power spectrum of faint galaxies, $C_{ff}$, are shown for a $0.1 \, \rm eV$ neutrino, $n_f=n_l=0.02~ h^3 {\rm Mpc}^{-3}$, and $b_l=2b_f=2$. } 
\label{noise_signal}
\end{center}
\end{figure}
The amplitude of our signal, ${\rm ~Im~} C_{fl}$, its error $\Delta {\rm ~Im~} C_{fl}$, and the auto-power spectrum of faint galaxies  is shown in Figure \ref{noise_signal}. This figure shows that the signal is dominated by large scale modes. The decline of the signal at small scales suggests that the measurement isn't improved by summing lots of modes. It is therefore not sample variance limited.

 \begin{figure*}
\begin{center}
\label{alpha_r_100spheres}
\includegraphics[width=8.5cm]{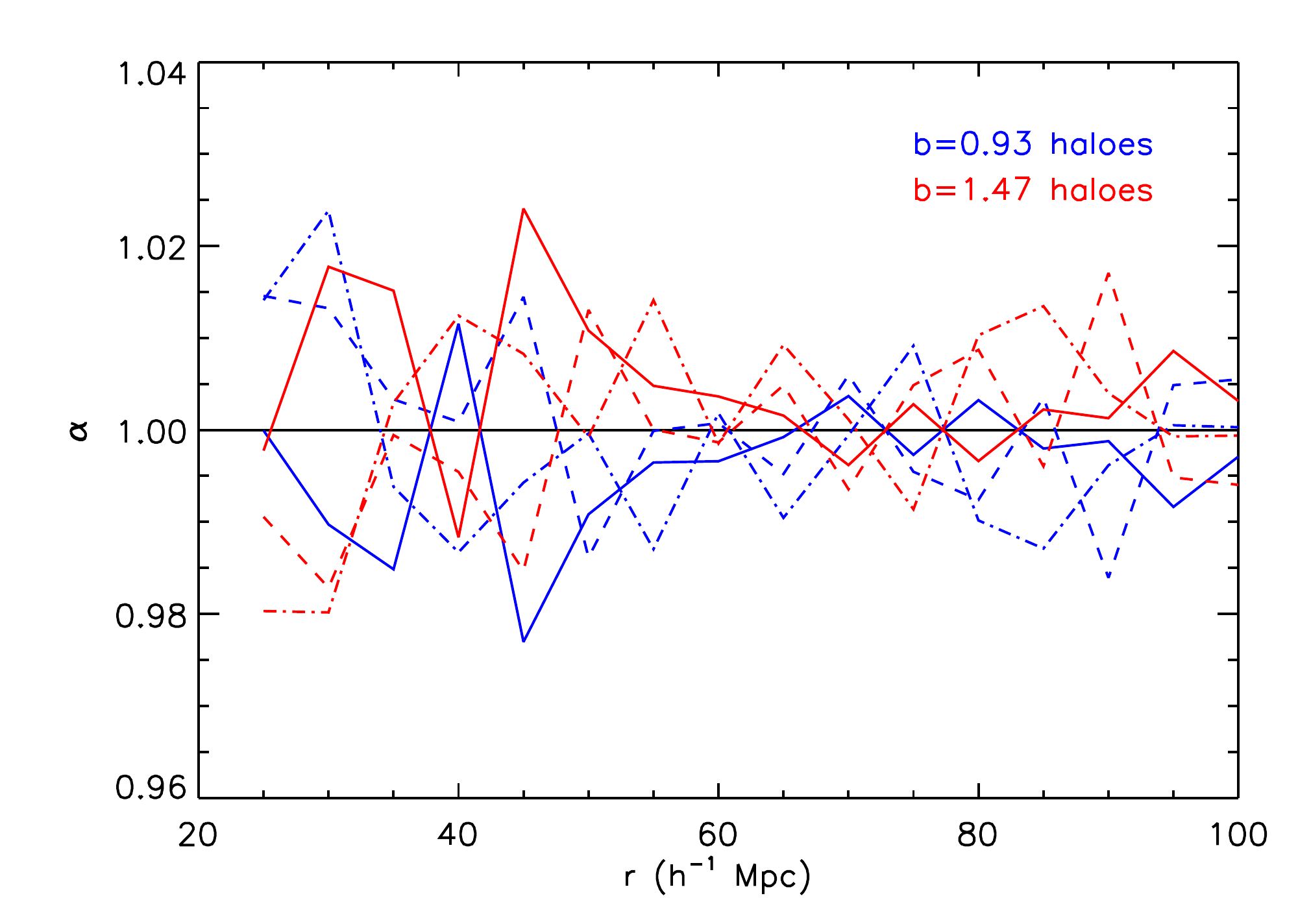}
\includegraphics[width=8.5cm]{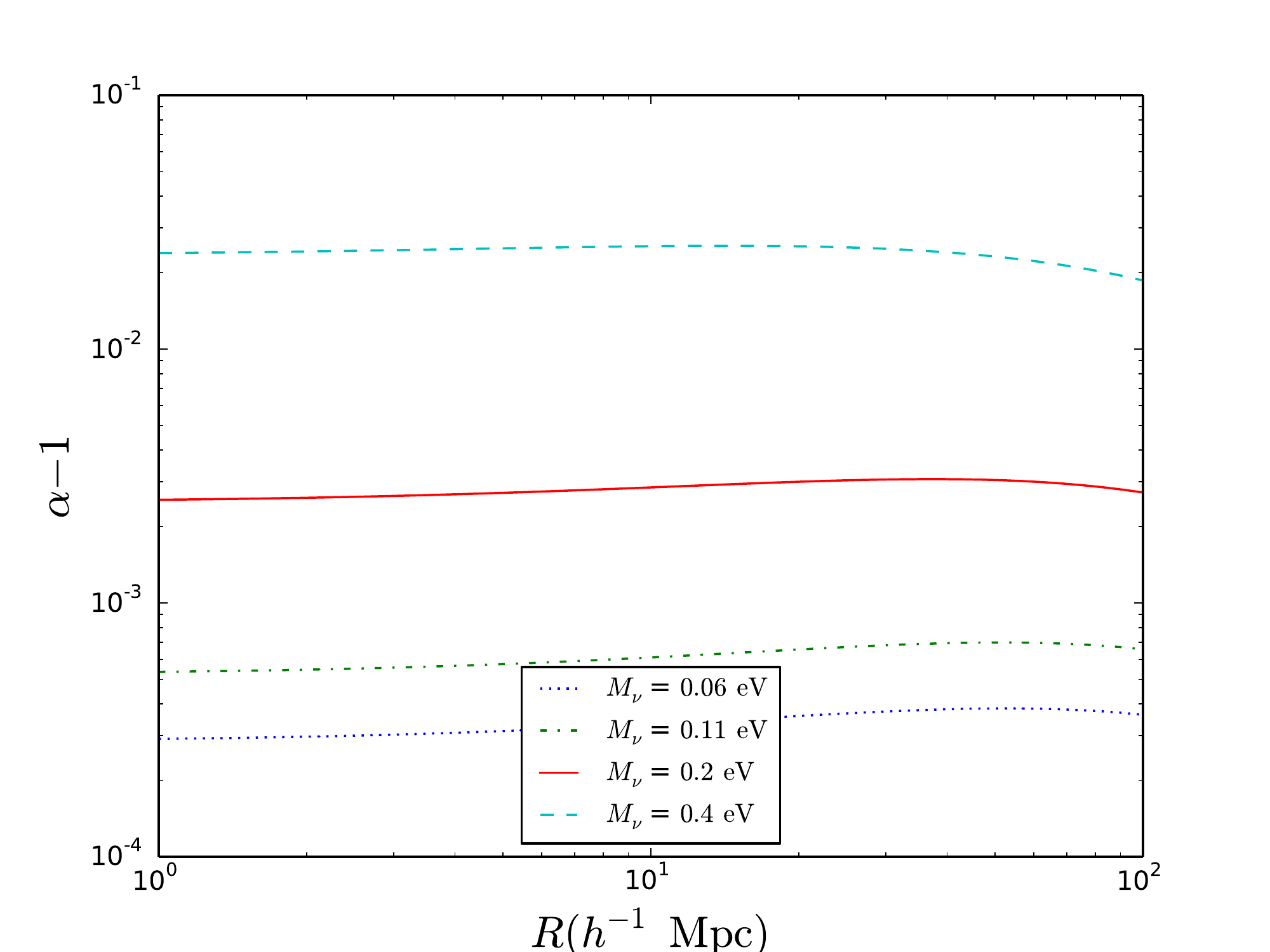}
\caption{(a)(Left) Fitted value of $\alpha$ over 100 spheres of radius $r$, for the $(x,y,z)$ components (solid, dashed and dot-dashed lines) of the bulk flow of our two halo samples. (b)(Right) $\Lambda$CDM prediction for $\alpha$ from Eq. (\ref{eq:alpha_r}) [for a $10^{15}(h^{-1} \msol$ halo)] as a function of top hat window function radius $R$ and different sum of neutrino mass. } 
\label{alpha}
\end{center}
\end{figure*}

To evaluate the signal-to-noise, we should also note that the neutrino-CDM relative velocity is not uniform, and vanishes on large scales (see Figure \ref{v_nuc_r}). To this end, we evaluate $\langle \dot{\phi}_{\bf k}^2 \rangle $, we note that only the coherent part of ${\bf v}_{\nu c}$ contributes to the phase shift, Eq. \ref{phase_shift}:
\begin{eqnarray}
\phi_{i \bf k} &\simeq& \frac{ \beta  a^2{\bf k \cdot v}_{i c}(<k,z)}{k^3}\nonumber \\
 &=&  \frac{\beta a^2 }{k^3} {\bf k}\cdot \int \frac{d^3k'}{(2\pi)^3} {\bf v}_{ic,{\bf k'}} \Theta(k-k')  \widetilde{W}(k'R)\nonumber \\
\end{eqnarray} where $\beta \equiv \frac{2G m^4_i \mu H}{\hbar^3}.$ 
This yields:
\begin{eqnarray}
\langle \dot{\phi}^2_{i \bf k} \rangle  &=&  \frac{\alpha H(a)}{k^2} \int^{k} \frac{dk'}{k'}  \left(3 a^3 \Delta_{vic}(k',z) - a^2 \frac{d \Delta_{vic}(k',z)}{dz}\right)^2 \nonumber \\
&& \widetilde{W}^2(k'R),
\end{eqnarray}
where $\Delta_{vic}$ was defined in Eq. (\ref{Delta_v}).

The total signal-to-noise squared for a single neutrino, $i$ for all modes is given by 
\begin{eqnarray}
\left(\frac{S}{N}\right)_i^2 &=& \sum_k \frac{|{\rm ~Im~} C_{fl} |^2}{ (\Delta {\rm ~Im~} C_{fl})^2},  \nonumber \\
&=& \frac{2 V_{\rm eff}}{(2\pi)^3} \int d^3k \left\langle \left|\frac{P_k \dot{\phi}_{i \bf k}\Delta b}{H}\frac{{ k_x^2}}{ |{\bf k}|^2}  \right|^2\right\rangle|\det C|^{-1}\nonumber \\
&=& \frac{2 V_{\rm eff}}{(2\pi^2)} \int^{\rm k_{\rm max}} d k k^2 P_k^2 \Delta b^2 \frac{\langle |\dot{\phi}_{i \bf k}^2| \rangle}{5H^2} |\det C|^{-1},\nonumber \\
\end{eqnarray}
where $\Delta b =  b_l - b_f.$ For our estimates, we have set $\mu = 1$ in $\dot{\phi}_{i \bf k}$ \textcolor{black}{(recall that $\mu$ ranges between 0.7 and 1 and the signal-to-noise is directly proportional to $\mu$)}. Given that the velocities of different neutrino species are highly correlated, the total signal-to-noise can be well-approximated by the sum of individual signal-to-noises, i.e.
\begin{equation}
\left(\frac{S}{N}\right)_{tot} \approx \sum_{i=1}^3 \left(\frac{S}{N}\right)_i.
\end{equation}

\section{Testing nonlinear effects using $N$-body simulations}
\label{sec:nonlinearities}

We shall next examine to what extent the nonlinear structure formation in standard $\Lambda$CDM model could mimic the effects of neutrino  dynamical friction on cosmological haloes.
We therefore use an $N$-body simulation without neutrinos in order to systematically account for the nonlinear effects.

The signal we study is the difference in relative displacement due to neutrinos, $\Delta x$, for two different tracers, given by  $(\Delta x_1 - \Delta x_2) = \frac{1}{2} (\Delta v_{\nu c , 1}-\Delta v_{\nu c,2}) t = \frac{1}{2} v_{\rm rel} t$, where we denote the relative velocity between tracers 1 and 2 by $v_{\rm rel}$. We want to know whether nonlinear velocities due to growth of structure on small scales could produce a signal that could interfere with the signal from massive neutrinos.

To this end, we use an $N$-body simulation without massive neutrinos to investigate the possibility of any existing contaminating signal from nonlinear effects. Assuming the neutrino streaming direction in any given volume is the same direction as the CDM bulk flow $\vc{v}_c$ (which it should be to first order), then we would expect the relative velocity $\vc{v}_{\rm rel}$ between halo populations 1 and 2 to be in that direction also. So we want to know whether there is any nonzero correlation between the direction of $\vc{v}_{\rm rel}$ and $\vc{v}_c$ in the absence of massive neutrinos. That is, we want
\begin{equation}
\langle \hat{\vc{v}}_{\rm rel}  \cdot \hat{\vc{v}}_c \rangle = 0,
\end{equation}
 where the averaging is over independent volumes, which we take to be spheres of radius $R$, and the hat denotes the unit vector.

This effect is tested using the Gigaparsec WiggleZ (GiggleZ) $N$-body simulation \citep{poole2014}. The GiggleZ main simulation contains $2160^3$ dark matter particles in a periodic box of side $1 h^{-1}$ Gpc. The particle mass is $7.5 \times 10^9 h^{-1}M_\odot$, which allows bound systems with  masses $\gtrsim 1.5 \times 10^{11} h^{-1} M_\odot$ to be resolved. The clustering bias $b$ of the haloes range from $\sim$ 1 to greater than 2. Halo finding for GiggleZ was performed using \textsc{subfind} \citep{springel2001}, which utilizes a friends-of-friends (FoF) algorithm to identify coherent overdensities of particles and a substructure analysis to determine bound overdensities within each FoF halo. The resulting \textsc{subfind} substructure catalogues are rank-ordered by their maximum circular velocity ($V_{\rm max,sub}$) as a proxy for halo size. 
\begin{eqnarray}
\vc{v}_1 &=& \alpha_1 \vc{v}_c + \vc{n}_1 \nonumber \\
\vc{v}_2 &=& \alpha_2 \vc{v}_c + \vc{n}_2.
\end{eqnarray}
In the absence of neutrinos, the equivalence principle implies that $\alpha \equiv 1$, so  we expect $\langle \vc{v}_{\rm rel} \rangle = 0$, i.e. $\alpha_1 = \alpha_2 = 1$, and $\langle \vc{v}_{\rm rel} \cdot \vc{v}_c \rangle = 0$.

In the presence of massive neutrinos, we expect that for haloes of mass $M$ within a volume of radius $R$,
\begin{equation}
\label{eq:alpha_r}
\alpha(R) = 1 + \frac{\langle \vc{v}_c(R) \cdot \Delta \vc{v}_{\nu c}(M,R) \rangle} {\langle \vc{v}_c(R) \cdot \vc{v}_c(R) \rangle}.
\end{equation}

Figure (\ref{alpha}a) shows the measured value of $\alpha$ from the relative velocity of our halo samples, which is consistent with zero at $< \%$ level, with no evident systematic bias. A larger sample and/or simulation box will lead to lower stochastic noise in $\alpha$, and can potentially reveal a systematic bias, albeit at a lower level. 

The expected value of $\alpha$ can be calculated from our equation for $\Delta v_{\nu c}$ and the $\Lambda$CDM prediction for $\vc{v}_c$. The 1D rms velocity dispersion of CDM is

\bea
\label{eqn_rms_velocity}
\langle v_c^2 \rangle &=& \int \frac{d^3k}{(2\pi)^3} |v_c(k)|^2 \widetilde{W}^2(kR) \nonumber \\
&=& a^2\int  dk \frac{\mathcal{P}_\chi}{k^3} \left| \dot{T}_c(k) \right|^2 \widetilde{W}^2(kR).
\eea

For the halo model estimation, $\Delta v_{\nu c}$ is given by Eq. (\ref{eq_deltav_halomodel})
and thus

\bea
&&\langle \vc{v}_c(R) \cdot \Delta \vc{v}_{\nu c}(m_h,R) \rangle = - \frac{2 \mu(\vc{v}_{\nu c},\vc{k}) G^2  M_{\rm h}}{ 3 \pi H \hbar^3 }\sum_{i=1}^3{m_i^4} \nonumber \\
&& \int \frac{dk}{k} \langle  \vc{v}_c \vc{v}_{i c} (<k) \rangle   \left[ \frac{}{}u(k|M_{\rm h})^2+ \frac{b(M_{\rm h}) P_{\rm CDM}(k) u(k|M_{\rm h})}{M_{\rm h}} \right.\nonumber \\
&& \left.   \int d M' \frac{dn}{d M'} b(M') M'  u(k|M') \frac{}{}  \right]
\eea
and 
\begin{equation}
\langle  \vc{v}_c \cdot \vc{v}_{i c} (<k) \rangle =  \int _0^k \frac{dk'}{k'} \Delta_{v c}(k',z) \Delta_{v i c}(k', z) \widetilde{W}^2(k'R) .
\end{equation}

The $k$ integral is over the Fourier modes of the halo distribution while the $k'$ integral is due to the modes from the neutrino distribution. We plot $\alpha$ as a function of top-hat window function radius $R$ and different sum of neutrino mass in Figure ~(\ref{alpha}b) for a halo of mass $M_h=10^{15}h^{-1}\msol$. Clearly, the predicted value of $\alpha-1$ is quite small. For example, for $M_\nu = 0.11\,\ev$, the value of $\alpha-1$ is below $10^{-3}$ for all radii requiring 0.1 percent-level precision of the bulk flow to detect the neutrino effects.

\label{lastpage}
\end{document}